\documentclass[preprint,aps,floatfix,nofootinbib,a4paper, superscriptaddress,prd]{revtex4-2}
\pdfoutput=1

\usepackage{amsmath, amssymb, amsfonts, amsthm, latexsym, epsfig, mathrsfs, xcolor, bbm, slashed, braket, cancel}

\usepackage[inline]{enumitem}

\usepackage{setspace}
\usepackage[marginal, multiple]{footmisc}

\usepackage[T1]{fontenc}
\usepackage[utf8]{inputenc}
\usepackage{lmodern}

\usepackage[colorlinks, allcolors=blue!70!black, linktocpage]{hyperref}

\makeatletter
\def\p@subsection{\thesection.}
\def\p@subsubsection{\thesection.\thesubsection.}
\makeatother 

\theoremstyle{plain}

\theoremstyle{definition}

\theoremstyle{remark}

\DeclareSymbolFont{matha}{OML}{txmi}{m}{it}
\DeclareMathSymbol{v}{\mathord}{matha}{118}

\begin{document}
\count\footins = 800 
\setstretch{1.2}


\title{Dynamical black hole entropy beyond general relativity from the Einstein frame}


\author{Delong Kong}
\email{kongdelong22@mails.ucas.ac.cn}
\affiliation{\textit{School of Physical Sciences, University of Chinese Academy of Sciences, Beijing 100049, China}}

\author{Yu Tian}
\email{ytian@ucas.ac.cn}
\affiliation{\textit{School of Physical Sciences, University of Chinese Academy of Sciences, Beijing 100049, China}}

\author{Hongbao Zhang}
\email{hongbaozhang@bnu.edu.cn}
\affiliation{School of Physics and Astronomy, Beijing Normal University, Beijing 100875, China}
\affiliation{Key Laboratory of Multiscale Spin Physics, Ministry of Education, Beijing Normal University, Beijing 100875, China}

\author{Jinan Zhao}
\email{jinanzhao@mail.bnu.edu.cn}
\affiliation{School of Physics and Astronomy, Beijing Normal University, Beijing 100875, China}
\affiliation{Key Laboratory of Multiscale Spin Physics, Ministry of Education, Beijing Normal University, Beijing 100875, China}

\begin{abstract}

Recently Hollands, Wald and Zhang proposed a new formula for the entropy of a dynamical black hole for an arbitrary theory of gravity obtained from a diffeomorphism covariant Lagrangian via the Noether charge method\cite{Hollands:2024vbe}. We present an alternative, pedagogical derivation of the dynamical black hole entropy for $f(R)$ gravity as well as canonical scalar-tensor theory by means of conformal transformations. First, in general relativity we generalize Visser and Yan's pedagogical proof of the non-stationary physical process first law to black holes that may not be in vacuum, and give a pedagogical derivation of the second-order behavior of the dynamical black hole entropy for vacuum perturbations by considering the second-order variation of the Raychaudhuri equation. Second, we apply the derivation for general relativity to theories in the Einstein frames, and then recast the conclusions in their original frames. We show that the dynamical black hole entropy formulas of these theories satisfy both the non-stationary physical process first law and the non-stationary comparison first law via the Einstein frame. We further study the second-order behavior of the dynamical black hole entropy for vacuum perturbations, and observe that the second law is obeyed at second order in those theories. Using the Einstein frame, we also determine the relationship between the dynamical black hole entropy and the Wald entropy of the generalized apparent horizon in the original frame.

\end{abstract}

\maketitle
\newpage

\tableofcontents


\section{Introduction}	

The discovery of the laws of black hole thermodynamics is one of the great achievements of fundamental physics\cite{Bardeen:1973gs,Bekenstein:1973ur,Hawking:1975vcx}. These laws combine gravity, quantum theory and thermodynamics within one stunning framework. Moreover, they provide some of the deepest insights on the fundamental nature of the quantum theory of gravity. For general relativity (GR), the black hole entropy $S$ is given by the Bekenstein-Hawking formula\cite{Bekenstein:1973ur,Hawking:1975vcx}
\begin{equation}
    S_{\text{BH}} = \frac{A}{4G},
\end{equation}
where $A$ is the area of the event horizon and $G$ is Newton's constant. While for an arbitrary theory of gravity obtained from a diffeomorphism covariant Lagrangian, the entropy of a stationary black hole at the bifurcation surface $\mathcal{B}$ is given by the Wald entropy, which is defined as the Noether charge and for the $f(\text{Riemann})$ theories of gravity it is written as\cite{Wald:1993nt,Iyer:1994ys}
\begin{equation}
    S_{\text{Wald}} = - 8\pi \int_{\mathcal{B}} \mathrm{d} A \ \frac{\partial L}{\partial R_{uvuv}},
\end{equation}
where $v$ is the future-directed affine parameter of the null generator of the future horizon, $u$ denotes the affine null distance away from the horizon and is also future-directed. Furthermore, Iyer and Wald proposed a formula $S_{\text{Iyer-Wald}}$ for dynamical black hole entropy on an arbitrary cross-section $\mathcal{C}$ of the horizon by expanding the Wald entropy in fields and their derivatives, and keeping the terms that depend only on boost-invariant fields\cite{Iyer:1994ys}.
However, the second law of black hole mechanics does not seem to hold for $S_{\text{Iyer-Wald}}$. And that proposed
dynamical entropy formula is not field-redefinition invariant. Moreover, the Noether charge method used to derive the black hole entropy is subjected to a number of ambiguities for non-stationary black holes, identified by Jacobson, Kang and Myers (JKM)\cite{Jacobson:1993vj}. Wall derived a second law for higher curvature gravity in a perturbative context, and resolved some ambiguities in Wald's Noether charge method\cite{Wall:2015raa}. For $f(\text{Riemann})$ gravity the Wall entropy is given by
\begin{equation}
    S_{\text{Wall}} = - 8\pi \int_{\mathcal{C}(v)} \mathrm{d} A \left(\frac{\partial L}{\partial R_{uvuv}} \right. \left.- 4 \frac{\partial^2 L}{\partial R_{uiuj} \partial R_{vkvl}} K_{ij(u)}K_{kl(v)}\right). 
\end{equation}
Where $K_{ij(a)}$ is the extrinsic curvature of the horizon in the $a$ direction. At the bifurcation surface both $S_{\text{Iyer-Wald}}$ and $S_{\text{Wall}}$ reduce to $S_{\text{Wald}}$, and they are equal to $S_{\text{BH}}$ for general relativity.  
 
Although the second law for $S_{\text{BH}}$ and the linearized second law for $S_{\text{Wall}}$ hold for non-stationary perturbations, this is not the case for the first law. The first law often does not hold for non-stationary perturbations of a stationary black hole, and if it does, the black hole entropy cannot be evaluated at an arbitrary cross-section of the event horizon of the perturbed non-stationary black hole\cite{Visser:2024pwz}. Recently, Hollands, Wald and Zhang proposed a strategy to overcome these two limitations\cite{Hollands:2024vbe}. They derived the first law by applying the Noether charge method to non-stationary perturbations of a stationary black hole background, and introduced a new program to the definition of dynamical black hole entropy valid to leading order, on the basis of the validity of a local, “physical process version” of the first law of black hole mechanics. Visser and Yan generalized and improved their work in a number of ways, and gave a more pedagogical proof of the physical process first law for black holes in general relativity\cite{Visser:2024pwz}.

Next let us review the key results of \cite{Hollands:2024vbe} for the dynamical black hole entropy and the non-stationary first law. For a non-stationary dynamical black hole, the entropy $S$ is no longer equal to the Bekenstein-Hawking entropy for GR or to the Wall entropy for higher curvature gravity. The formula for dynamical black holes differs from the usual Noether charge formula by a nontrivial dynamical correction term. For general relativity, the dynamical black hole entropy was defined as
\begin{equation}
    S_{\text{dyn}} = \left(1-v\frac{\mathrm{d}}{\mathrm{d} v}\right)S_{\text{BH}},
\end{equation}
where $v$ is the future-directed affine null parameter along the future horizon, and it is equal to 0 at the bifurcation surface $\mathcal{B}$. Notably, the derivative term is invariant under the scaling transformation of the affine parameter $v \to a(x^i) v$, where $x^i$ are codimension-2 spatial coordinates on the horizon. It was also shown that, to leading order in perturbation theory, the formula of dynamical black hole entropy for general relativity is equal to the Bekenstein-Hawking entropy of the apparent horizon. Furthermore, the dynamical black hole entropy was generalized to higher curvature gravity as\cite{Hollands:2024vbe}
\begin{equation}
    S_{\text{dyn}} = \left(1-v\frac{\mathrm{d}}{\mathrm{d} v}\right)S_{\text{Wall}}.
\end{equation}
	
By applying the Noether charge method to non-stationary perturbations, Hollands, Wald and Zhang have shown that their formula for dynamical black hole entropy satisfies both the non-stationary comparison first law and the non-stationary physical process first law. The non-stationary comparison version of the first law for an arbitrary horizon cross-section compares two vacuum black hole geometries 
\begin{equation}
    \frac{\kappa}{2 \pi} \delta S_{\text{dyn}}\left[\mathcal{C}(v)\right] = \delta M - \Omega_\mathcal{H} \delta J,
\end{equation}
where $M$ and $J$ are the mass and angular momentum of the black hole, respectively. $\Omega_\mathcal{H}$ is the
angular velocity of the horizon. And $\kappa$ denotes the surface gravity of the black hole. For vacuum perturbations of a stationary black hole, the dynamical entropy is “time independent” to first order, i.e., $\delta S_{\text{dyn}} [\mathcal{C}(v)] = \delta S_{\text{dyn}}[\mathcal{B}]$, where $\mathcal{C}(v)$ is an arbitrary cross-section. In order to study the nontrivial time variation of black hole entropy, we may unseal an external stress-energy, $\delta T_{ab}$, in the first-order perturbation. For perturbations sourced by external matter fields, the non-stationary physical process first law reads\footnote{Note that in the traditional treatments of the physical process first law, the black hole starts and ends in a stationary state\cite{Gao:2001ut,Poisson:2009pwt}.}
\begin{equation}
    \begin{split}
        \frac{\kappa}{2 \pi} \Delta \delta S_{\text{dyn}} &=\int_{v_1}^{v_2} \mathrm{d}v \int_{\mathcal{C}(v)} \mathrm{d} A \
		\kappa v \delta T_{vv} = \Delta \delta M - \Omega_\mathcal{H} \Delta \delta J,
    \end{split}
\end{equation}
where $\Delta$ denotes the difference between two cross-sections $\mathcal{C}(v_1)$ and $\mathcal{C}(v_2)$, and $\delta$ stands for the first-order perturbation around the stationary background. As a corollary of the physical process first law, the second law holds for the dynamical black hole entropy to first order as long as the stress-energy tensor satisfies the null energy condition $\delta T_{vv} \ge 0$. While for vacuum perturbations we have to pay attention to second order in perturbation theory to study the leading-order change of the dynamical black hole entropy. And for vacuum perturbations the second law is obeyed at second order in general relativity since the “modified canonical energy flux” is positive for GR. However, in more general theories of gravity the “modified canonical energy flux” is not necessarily positive, and the second law presumably would not hold in more general theories of gravity for vacuum perturbations.

In this paper we present an alternative, pedagogical proof of both the non-stationary physical process first law and the non-stationary comparison first law for the $f(R)$ gravity as well as the canonical scalar-tensor theory via conformal transformations. To begin with, we generalize Visser and Yan's pedagogical derivation of the physical process first law for GR to black holes that may not be in vacuum. We also present a pedagogical derivation of the second-order behavior of $S_{\text{dyn}}$ for vacuum perturbations in GR by studying the second variation of the Raychaudhuri equation. Since conformal transformations preserve the causal structures of spacetimes, and the conformal factors that put theories into the Einstein frames are independent of the Killing time $\tau$ in the stationary spacetimes, the spacetimes given by conformal transformations are stationary black hole solutions to the Einstein equation. We are capable of applying the proof of the first law for GR to theories in the Einstein frames, and then recast them in their original frames. As the perturbations are non-stationary, the gauge conditions for perturbations that fix the affine parameter $v^E$ in the Einstein frame will not necessarily fix the affine parameter $v$ in the original frame. However, those gauge conditions do not restrict the way we perturb the stationary black hole background. Instead, they tell us how to identify spacetime points in the two slightly different spacetimes. We show that the switching from $v^E-$identification to $v-$identification leads to second-order corrections, and the first laws are invariant to first order under the reidentification of spacetime points. For vacuum perturbations we further study the second-order behavior of the dynamical black hole entropy, and observe that the second law holds at second order both in the Einstein frames and in the original frames. Using the Einstein frame we also determine the relationship between the dynamical black hole entropy of the cross-section and the Wald entropy of the generalized apparent horizon in the original frame, which was not given before.

The rest of the paper is organized as follows. In Sec. \ref{sec2} we introduce the geometric setup and impose gauge conditions on perturbations.
In Sec. \ref{sec3} we generalize Visser and Yan's pedagogical proof of the physical process first law to non-vacuum solutions, and study the second-order behavior of the dynamical black hole entropy for vacuum perturbations in GR.
In Secs. \ref{sec4} and \ref{sec5} we present the proof of both the physical process first law and the comparison first law for $f(R)$ gravity and canonical scalar-tensor theory, study the second-order behavior of the entropy for vacuum perturbations in those theories, and determine the relationship between the dynamical black hole entropy and the Wald entropy of the generalized apparent horizon in the original frame.
Sec. \ref{sec6} is devoted to the summary of the paper and some discussions.
In Appendix \ref{appendix} we calculate the modified canonical energy flux for $f(R)$ gravity via the covariant phase space formalism, and give a nontrivial check of the second-order behavior of $S_{\text{dyn}}$ for vacuum perturbations.

We will mainly follow the notation and conventions of \cite{Wald:1984rg}. Throughout this paper, we set $c = \hbar = k_B = 1$ while keep Newton's constant $G_D$ explicit.

	
\section{Stationary black hole background and gauge conditions for perturbations}\label{sec2}
	
In this section we introduce the geometry of the stationary black hole background and impose gauge conditions on non-stationary perturbations used to study the first law. Our geometric setup mainly follows that of \cite{Visser:2024pwz}.
\begin{figure}
    \centering
    \includegraphics[width=.5\linewidth]{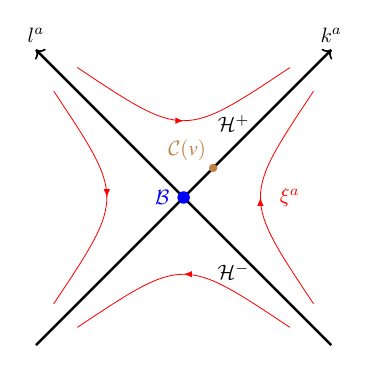}
    \caption{The stationary black hole background.}
    \label{fig:Stationary_BH}
\end{figure}

Consider a $D-$dimensional stationary black hole background geometry $(M,g_{ab})$, as shown in FIG. \ref{fig:Stationary_BH}. We assume that the black hole spacetime is asymptotically flat and electrically neutral. The event horizon of this black hole coincides with the bifurcate Killing Horizon $\mathcal{H}$. We label the future horizon by $\mathcal{H}^+$, the past horizon by $\mathcal{H}^-$, and the bifurcation surface by $\mathcal{B}$. The Killing field normal to the Killing horizon is denoted by $\xi^a$, and it is a Killing symmetry of the metric $g_{ab}$ as well as the matter fields $\phi$. i.e.,
\begin{equation}
    \mathcal{L}_\xi g_{ab} = 0, \quad \mathcal{L}_\xi \phi = 0.
\end{equation}
	
We are mainly interested in the part of $\mathcal{H}^+$ that lies to the future of $\mathcal{B}$. We can always erect a set of null zweibein bases $(k^a,l^a)$ on $\mathcal{H}^+$, where $k^a$ is the future directed null normal to $\mathcal{H}^+$ and it is affinely parameterized as $k^a \overset{\mathcal{H}^+}{=} (\partial_v)^a$ with $v$ being set to $0$ on $\mathcal{B}$, $l^a$ is an (future-directed) ingoing auxiliary null vector field and $k^a l_a = -1$ holds on the horizon. We can also extend $l^a$ off the horizon by solving $l^b \nabla_b l^a = 0$ and denote the affine null distance away from $\mathcal{H}^+$ by $u$ to identify $l^a = (\partial_u)^a$. Finally, we can extend $k^a$ off the horizon such that it commutes with $l^a$, i.e., $\left[k,l\right]^a = 0$. Since $\xi^b \nabla_b \xi^a \overset{\mathcal{H}^+}{=} \kappa \xi^a$ and $k^b \nabla_b k^a \overset{\mathcal{H}^+}{=} 0$, we can scale $k^a$ properly such that 
\begin{equation}
    \xi^a = (\partial_\tau)^a \overset{\mathcal{H}^+}{=} C e^{\kappa \tau} k^a = \kappa v k^a,
\end{equation}
where $C$ is a constant. We can also decompose the metric on $\mathcal{H}^+$ as
\begin{equation}
    g_{ab} \overset{\mathcal{H}^+}{=}-k_a l_b - l_a k_b + \gamma_{ab},
\end{equation}
where $\gamma_{ab}$ is the intrinsic codimension-2 spatial metric of the cross-section $\mathcal{C}(v)$ satisfying $\gamma_{ab} = \gamma_{(ab)}$ and $\gamma_{ab} k^a \overset{\mathcal{H}^+}{=} \gamma_{ab}l^a \overset{\mathcal{H}^+}{=} 0$.
	
We would like to perturb the stationary black hole background $g_{ab} \to g_{ab} + \delta g_{ab}$ and the matter fields $\phi \to \phi + \delta \phi$ to study the first law for non-stationary variations, where $\delta g_{ab}, \ \delta \phi \sim \mathcal{O} (\epsilon)$ are of first order in the perturbation. Since the stationary black hole spacetime and the perturbed spacetime are different, there exists certain gauge freedoms about how we should establish the coordinate systems in the perturbed geometry and how to identify spacetime points in the two slightly different spacetimes. In order to simplify the derivation of the first law, we impose the following gauge conditions on perturbations\cite{Visser:2024pwz}:
	
\begin{enumerate}
    \item The event horizon of the perturbed black hole is identified with the Killing horizon of the background geometry. And $\mathcal{H}^+$ is still described by $u=0$ and  $\mathcal{H}^-$ described by $v=0$ after the perturbation.  
    \item $k^a$ and $l^a$ are fixed under the perturbation
    \begin{equation}
		\delta k^a = 0, \quad \delta l^a = 0.
    \end{equation}
    and $k^a$ remains null normal to $\mathcal{H}^+$ and $l^a$ remains null everywhere under the perturbation. Combined with the requirement $\delta (k^a l_a) = 0$ this implies the following conditions
    \begin{equation}
		k^a \delta g_{ab} \overset{\mathcal{H}^+}{=} 0, \quad
		l^a \delta g_{ab} = 0.
    \end{equation}
    We further require that $k^a$ is still affinely parameterized on $\mathcal{H}^+$ and $l^a$ affinely parameterized everywhere after the perturbation
    \begin{equation}
		\delta (k^b \nabla_b k^a) \overset{\mathcal{H}^+}{=} 0, \quad
		\delta (l^b \nabla_b l^a) = 0.
    \end{equation}
    \item The Killing vector field $\xi^a$ remains null and tangent to the geodesic generators of the perturbed black hole. Combined with $\xi^a \delta g_{ab} \overset{\mathcal{H}}{=} 0$ this means that $\delta \xi^a$ is proportional to $k^a$ on $\mathcal{H}^+$ and to $l^a$ on $\mathcal{H}^-$. 
		
\end{enumerate}
The gauge condition 1 simply means that we compare the event horizon of the perturbed black hole and the Killing horizon of the stationary background. Condition 2 tells us how to establish the $u,v$ coordinate systems in the perturbed spacetimes. Moreover, we identify points with identical coordinates in the two slightly different spacetimes. These conditions do not mean that the Killing field should be fixed under the perturbation. If we allow the surface gravity $\kappa$ to vary, $\xi^a$ will change in the perturbed geometry.

	
\section{Dynamical black hole entropy in general relativity}\label{sec3}
	
In this section, we present the formula for dynamical black hole entropy based on the non-stationary physical process first law in GR. This derivation is completed by means of integrating the linearized Raychaudhuri equation on the horizon between two arbitrary cross-sections. A similar derivation was given in \cite{Visser:2024pwz} for black hole solutions to the vacuum Einstein equation. While our derivation is less restrictive since we allow the metric to couple to matter fields via the Einstein equation in the stationary background. For vacuum perturbations we also provide a pedagogical derivation of the second-order behavior of the dynamical black hole entropy by considering the second-order variation of the Raychaudhuri equation. We also note that a similar deduction for vacuum perturbations was established in \cite{Rignon-Bret:2023fjq} before us.
	
Let us consider the stationary black hole background and the perturbed geometry introduced in the last section. The Raychaudhuri equation for the congruence of null geodesics on the future horizon $\mathcal{H}^+$ reads
\begin{equation}\label{Ray}
    \frac{\mathrm{d} \theta}{\mathrm{d} v} = - \frac{1}{D-2} \theta ^2
    - \sigma_{ab} \sigma^{ab} + \omega_{ab} \omega^{ab} - R_{ab} k^a k^b.
\end{equation}
As the black hole background is stationary, the expansion $\theta$ and shear tensor $\sigma_{ab}$ vanish in the unperturbed spacetime geometry. What's more, the rotation tensor $\omega_{ab}$ is equal to 0 since $k^a$ is hypersurface orthogonal. With the aid of the Einstein equation, the last term of (\ref{Ray}) can be written as
\begin{equation}
    R_{ab} k^a k^b = 8 \pi G_D \left( T_{ab} - \frac{1}{2} T g_{ab} \right) k^a k^b = 8\pi G_D T_{ab} k^a k^b. 
\end{equation}
Where $G_D$ is the $D-$dimensional Newton's constant. As a corollary of the Raychaudhuri equation combined with the Einstein equation, $T_{ab} k^a k^b$ is equal to zero on the horizon in the stationary geometry. Otherwise the last term on the right-hand side of (\ref{Ray}) would not be zero\cite{Poisson:2009pwt}. Physically this condition can be interpreted as matter cannot be flowing across the event horizon. We multiply the Raychaudhuri equation on both sides by $\kappa v$, integrate it over the future horizon between cross-sections $\mathcal{C}(v_1)$ and $\mathcal{C}(v_2)$, and then vary this equation. Since $\frac{\mathrm{d} \theta}{\mathrm{d} v}$ and $T_{ab} k^a k^b$ vanish in the unperturbed geometry, $\delta$ acts only on $\theta$ on the left-hand side and only on $R_{ab}$ on the right-hand side, and the right-hand side is left with $8 \pi G_D \kappa v \delta T_{ab} k^a k^b$ due to our gauge condition $\delta g_{ab} k^a k^b = 0$. Recalling that $\xi^a = \kappa v k^a$ on the future horizon, to first order the result is
\begin{equation}\label{int}
    \kappa \int_{v_1}^{v_2} \mathrm{d}v \int_{\mathcal{C}(v)} \mathrm{d} A \ v \frac{\mathrm{d} \delta \theta}{\mathrm{d} v}
    = - 8 \pi G_D \int_{v_1}^{v_2} \mathrm{d}v \int_{\mathcal{C}(v)} \mathrm{d} A \ \delta T_{ab} \xi^a k^b,
\end{equation}
where $\mathrm{d} A = \mathrm{d}^{D-2}x \sqrt{\gamma(x,v)}$ is the area element of the cross-section $\mathcal{C}(v)$. Let $x = (x^1,\cdots x^{D-2})$ be a coordinate system on $\mathcal{C}(v_1)$, we can extend this coordinate system to an arbitrary cross-section by requiring that all points on each null geodesic share the same values of $(x^1,\cdots x^{D-2})$. To exchange the order of integrals in (\ref{int}), define the function $\mathcal{A}(x,v)$ on $\mathcal{H}^+$ as\cite{Flanagan:1999jp}
\begin{equation}
    \mathcal{A}(x,v) = \exp \left[\int_{v_1}^v \mathrm{d} \tilde{v} \ \theta (x,\tilde{v})\right] = \frac{\sqrt{\gamma(x,v)}}{\sqrt{\gamma(x,v_1)}}.
\end{equation}
The integral on the left-hand side of (\ref{int}) can then be split into
\begin{equation}
    \begin{split}
        \int_{v_1}^{v_2} \mathrm{d}v \int_{\mathcal{C}(v)} \mathrm{d} A \
		v \frac{\mathrm{d} \delta \theta}{\mathrm{d} v} &= \int_{v_1}^{v_2} \mathrm{d}v \int_{\mathcal{C}} \mathrm{d}^{D-2}x \sqrt{\gamma(x,v)} v \frac{\mathrm{d} \delta \theta}{\mathrm{d} v}
        \\
        &= \int_{v_1}^{v_2} \mathrm{d}v \int_{\mathcal{C}} \mathrm{d}^{D-2}x \sqrt{\gamma(x,v_1)} \frac{\sqrt{\gamma(x,v)}}{\sqrt{\gamma(x,v_1)}} v \frac{\mathrm{d} \delta \theta}{\mathrm{d} v}\\
        &=  \int_{\mathcal{C}} \mathrm{d}^{D-2}x \sqrt{\gamma(x,v_1)} \int_{v_1}^{v_2} \mathrm{d}v \frac{\sqrt{\gamma(x,v)}}{\sqrt{\gamma(x,v_1)}} v \frac{\mathrm{d} \delta \theta}{\mathrm{d} v} \\
		&=\int_{\mathcal{C}(v_1)} \mathrm{d} A \int_{v_1}^{v_2} \mathrm{d}v \
		\mathcal{A}(x,v) v \frac{\mathrm{d} \delta \theta}{\mathrm{d} v},	
    \end{split}	
\end{equation}
and it can be integrated by parts
\begin{equation}
    \int_{\mathcal{C}(v_1)} \mathrm{d} A \int_{v_1}^{v_2} \mathrm{d}v \ v \frac{\mathrm{d} \delta \theta}{\mathrm{d} v} 
    = \int_{\mathcal{C}(v_1)} \mathrm{d} A \left[v \delta \theta\right]_{v_1}^{v_2} - \int_{\mathcal{C}(v_1)} \mathrm{d} A \int_{v_1}^{v_2} \mathrm{d}v \ \delta \theta,
\end{equation}
where we have used the fact that $\mathcal{A}(x,v) = 1$ in the stationary background. On the right-hand side of this equation, we may pull the variation $\delta$ to the front of the integrals as $\theta$ vanishes on the horizon of the unperturbed black hole background. And the second term on the right-hand side is 
\begin{equation}
    \begin{split}
		& \delta \int_{\mathcal{C}(v_1)} \mathrm{d} A \int_{v_1}^{v_2} \mathrm{d}v \  \theta \\
		= &\delta \int_{\mathcal{C}(v_1)} \mathrm{d} A \int_{v_1}^{v_2} \mathrm{d}v \ \mathcal{A}(x,v) \partial_v \ln \sqrt{\gamma(x,v)}  \\
		= &\delta \int_{\mathcal{C}(v_1)} \mathrm{d}^{D-2} x \sqrt{\gamma(x,v_1)}  \int_{v_1}^{v_2} \mathrm{d}v \ 
		\frac{\sqrt{\gamma(x,v)}}{\sqrt{\gamma(x,v_1)}} 
		\frac{\partial_v \sqrt{\gamma(x,v)}}{\sqrt{\gamma(x,v)}}  \\
		= &\delta \int_{\mathcal{C}(v_1)} \mathrm{d}^{D-2} x \left[\sqrt{\gamma(x,v_2)} - \sqrt{\gamma(x,v_1)}\right]  \\
		= &\Delta \delta A.
    \end{split}
\end{equation}
As a result, the left part of (\ref{int}) reads
\begin{equation}
    \kappa \int_{v_1}^{v_2} \mathrm{d}v \int_{\mathcal{C}(v)} \mathrm{d} A \
    v \frac{\mathrm{d} \delta \theta}{\mathrm{d} v} = 
    - \kappa \Delta \delta \left(\int_{\mathcal{C}(v)} \mathrm{d} A \
    (1- v \theta)\right).
\end{equation}
Thus we obtain the non-stationary physical process first law between two arbitrary cross-sections for GR
\begin{equation}\label{PPFL}
    \frac{\kappa}{2\pi} \Delta \delta S_{\text{dyn}} =  
    \int_{v_1}^{v_2} \mathrm{d}v \int_{\mathcal{C}(v)} \mathrm{d} A \
    \delta T_{ab} \xi^a k^b,
\end{equation}
where $S_{\text{dyn}}$ is the dynamical black hole entropy of the cross-section of the horizon
\begin{equation}
    S_{\text{dyn}}[\mathcal{C}] = \frac{1}{4G_D} \int_{\mathcal{C}(v)} \mathrm{d} A \left(1 - v \theta\right) = \left(1 - v \frac{\mathrm{d}}{\mathrm{d} v}\right) S_{\text{BH}}.
\end{equation}
The matter Killing energy flux between two cross-sections $\mathcal{C}(v_1)$ and $\mathcal{C}(v_2)$, relative to the Killing field $\xi^a$, is defined as
\begin{equation}
    \Delta E = \int_{v_1}^{v_2} \mathrm{d} v \int_{\mathcal{C}(v)} \mathrm{d} A \ T_{ab} \xi^a k^b.
\end{equation}
And $\Delta$ denotes the difference between $\mathcal{C}(v_1)$ and $\mathcal{C}(v_2)$. By the argument similar to (\ref{int}), if we vary $\Delta E$, $\delta$ acts only on $T_{ab}$. It was shown by Hawking that if a black hole is stationary, then it must be either static or axisymmetric\cite{Hawking:1971vc}. If the horizon Killing field is normalized as $\xi^a = \left(\partial_t\right)^a + \Omega_\mathcal{H} \left(\partial_\theta\right)^a$, where $\left(\partial_t\right)^a$ represents time translations at infinity, $\left(\partial_\theta\right)^a$ represents the rotational Killing vector and $\Omega_\mathcal{H}$ is the angular velocity of the black hole, then the mass and angular momentum transferred across the horizon are\cite{Poisson:2009pwt}
\begin{equation}
    \begin{split}
        \Delta \delta M_H &= \int_{v_1}^{v_2} \mathrm{d}v \int_{\mathcal{C}(v)} \mathrm{d} A \
		\delta T_{ab} \left(\partial_t\right)^a k^b,  \\
		\Delta \delta J_H &= -\int_{v_1}^{v_2} \mathrm{d}v \int_{\mathcal{C}(v)} \mathrm{d} A \
		\delta T_{ab} \left(\partial_\theta\right)^a k^b.
	\end{split}
\end{equation}
The matter Killing energy flux is then related to the change in mass and angular momentum of the black hole as\footnote{We emphasize that the mass and the angular momentum of the black hole are not equal to those of the spacetime if the black hole is not in vacuum. The matter distribution outside the black hole also contributes to the total mass and total angular momentum of the spacetime.}
\begin{equation}
    \int_{v_1}^{v_2} \mathrm{d}v \int_{\mathcal{C}(v)} \mathrm{d} A \
    \delta T_{ab} \xi^a k^b = 
    \Delta \delta M_H - \Omega_\mathcal{H} \Delta \delta J_H.
\end{equation}
Therefore, the physical process first law reads
\begin{equation}
    \frac{\kappa}{2\pi} \Delta \delta S_{\text{dyn}} =  
    \Delta \delta M_H - \Omega_\mathcal{H} \Delta \delta J_H.
\end{equation}
	
As the first corollary of the physical process first law, if the perturbation of the stress-energy tensor satisfies the null energy condition $\delta T_{ab} k^a k^b \ge 0$, for first-order perturbations sourced by external matter fields, the “linearized” second law holds for the dynamical black hole entropy
\begin{equation}
    \Delta \delta S_{\text{dyn}} \ge 0.
\end{equation}
And as the second corollary, for source-free perturbations of a stationary black hole, the comparison first law holds between an arbitrary cross-section $\mathcal{C}(v)$ and the spatial infinity $i_0$
\begin{equation}\label{CFL}
    \frac{\kappa}{2 \pi} \delta S_{\text{dyn}}[\mathcal{C}(v)] = \delta M - \Omega_\mathcal{H} \delta J,
\end{equation} 
where $M$ and $J$ are ADM definitions of the mass and angular momentum of the spacetime, respectively. This is because $\Delta \delta S_{\text{dyn}} = 0$ for source-free perturbations $\delta T_{ab} = 0$, $S_{\text{dyn}}$ equals the usual Bekenstein-Hawking entropy at the bifurcation surface, and it has been proven that Bekenstein-Hawking entropy satisfies the comparison first law at the bifurcation surface. It is noteworthy that this relation still holds if matter fields are present as long as the combined Einstein-matter system admits a Hamiltonian formulation\cite{Wald:1995yp,Sudarsky:1992ty}. 
	
In the case of vacuum perturbations there are no external matter fields and $\delta T_{ab} = 0$. Then (\ref{PPFL}) indicates that the dynamical black hole entropy is a constant at first order in perturbation theory. So we have to keep track of the second order in perturbation theory to study the leading-order behavior of the dynamical black hole entropy. Hollands, Wald and Zhang obtained the leading-order behavior in that case based on the varied fundamental identity of the covariant phase space formalism\cite{Hollands:2024vbe}. In what follows we are going to present a different derivation of the second-order behavior of $S_{\text{dyn}}$ by considering the second variation of the Raychaudhuri equation\cite{Rignon-Bret:2023fjq}.
	
For vacuum perturbations, the first-order variation of (\ref{Ray}) is
\begin{equation}
    \frac{\mathrm{d}}{\mathrm{d} v} \delta \theta =\delta R_{ab} k^a k^b = 0.
\end{equation}
Thus the perturbed expansion $\delta \theta$ is a constant along the null generators of $\mathcal{H}^+$. Recalling that $\theta$ vanishes at future infinity in the perturbed spacetime due to the teleological definition of the event horizon, the perturbed expansion $\delta \theta$ must vanish on $\mathcal{H}^+$\cite{Hollands:2012sf}. We then consider the second-order variation of (\ref{Ray}) to obtain
\begin{equation}
    \frac{\mathrm{d}}{\mathrm{d} v} \delta^2 \theta = 
		- 2 \delta \sigma_{ab} \delta \sigma^{ab}.
\end{equation}
We multiply this equation on both sides by $\kappa v$, integrate over the future horizon between $\mathcal{C}(v_1)$ and $\mathcal{C}(v_2)$, and integrate the left-hand side by parts in the same way shown above
\begin{equation}
    \kappa \int_{\mathcal{C}(v_1)} \mathrm{d} A \left[v \delta^2 \theta\right]_{v_1}^{v_2} 
    - \kappa \int_{\mathcal{C}(v_1)} \mathrm{d} A \int_{v_1}^{v_2} \mathrm{d}v \ \delta^2 \theta  
    - 2 \int_{v_1}^{v_2} \mathrm{d}v \int_{\mathcal{C}(v)} \mathrm{d} A \
    \kappa v \left(\delta\sigma_{ab}\delta \sigma^{ab}\right).
\end{equation}
What's more, on the left-hand side of this equation we may pull $\delta^2$ to the front of the integrals as both $\theta$ and $\delta \theta$ vanish. Finally the result is
\begin{equation}
    \frac{\kappa}{2\pi} \Delta \delta^2 S_{\text{dyn}} =  
    \frac{1}{4 \pi G_D} \int_{v_1}^{v_2} \mathrm{d}v \int_{\mathcal{C}(v)} \mathrm{d} A \
    \kappa v \left(\delta \sigma_{ab}\delta \sigma^{ab}\right).
\end{equation}
Our result agrees with that derived by varying the fundamental identity of the covariant phase space formalism\cite{Hollands:2024vbe}. The term $\delta \sigma_{ab} \delta \sigma^{ab}$ can be interpreted as the energy flux of the weak gravitational waves\cite{Ashtekar:2021kqj}. Since $\delta \sigma_{ab} \delta \sigma^{ab} \ge 0$, the second law holds at second order for vacuum perturbations of GR. 
	

\section{Dynamical black hole entropy in \texorpdfstring{$f(R)$}{f(R)} gravity}\label{sec4}
	
In this section we derive both the non-stationary physical process first law and the non-stationary comparison first law for $f(R)$ gravity using the Einstein frame. Our results and the dynamical black hole entropy formula for $f(R)$ gravity agree with those derived by means of the Noether charge method. For vacuum perturbations we also study the second-order behavior of the dynamical black hole entropy, and observe that the second law is obeyed both in the Einstein frame and in the original frame. What's more, we determine the relationship between the dynamical black hole entropy of the cross-section $\mathcal{C}$ and the Wald entropy of the generalized apparent horizon $\mathcal{T}$ in the original frame.
	
\subsection{The conformal transformation and the Einstein frame}
	
The $f(R)$ gravity is described by the action
\begin{equation}\label{f(R)}
	I_f = \frac{1}{16 \pi G_D} \int \mathrm{d}^D x \sqrt{-g} f(R).
\end{equation}
And in this paper, we assume that\footnote{We impose $f'(R) > 0$ to ensure that the effective gravitational coupling strength $G_{\text{eff}} = G / f'(R)$ for the $f(R)$ theory is positive\cite{Sotiriou:2008rp,Nojiri:2017ncd}. On the other hand, if $f''(R) = 0$, then $f(R)$ is linear to $R$, and the resulting theory is reduced to GR.} 
\begin{equation}
    f'(R) >0, \quad f''(R) \ne 0.
\end{equation}
We invent an auxiliary scalar field $\varphi$ to write the action (\ref{f(R)}) in an equivalent form
\begin{equation}\label{equ}
    \tilde{I}_f = \frac{1}{16 \pi G_D} \int \mathrm{d}^D x \sqrt{-g}\left[f'(\varphi)R + f(\varphi) - \varphi f'(\varphi)\right],
\end{equation}
because with the on-shell condition $\varphi = R$, this action reduces to the original one.
To utilize the conclusions of the last section, we turn to the Einstein frame and redefine the scalar field as follows\cite{Barrow:1988xh,Whitt:1984pd}
\begin{equation}\label{conformal transformation}
    \begin{split}
        g^E_{ab} \equiv \left[f'(\varphi)\right]^{\frac{2}{D-2}} g_{ab}, \quad
	\phi \equiv \frac{1}{\sqrt{16 \pi G_D}} 
        \sqrt{\frac{2\left(D-1\right)}{D-2}} \ln f'(\varphi).
        \end{split}
\end{equation}
As a result, the equivalent action (\ref{equ}) can be rewritten as
\begin{equation}\label{f(R)Ein}
    \tilde{I}_f = \int \mathrm{d}^D x \sqrt{-g^E}
    \left[\frac{R^E}{16 \pi G_D} - \frac{1}{2} g_E^{ab} \partial_a \phi \partial_b \phi - V(\phi) \right],
\end{equation}
where
\begin{equation}
    V(\phi) \equiv \frac{1}{16\pi G_D \left[f'(\varphi)\right]^{\frac{D}{D-2}}}
    \left( \varphi f'(\varphi) - f(\varphi) \right).
\end{equation}
Obviously, the action (\ref{f(R)Ein}) is the Einstein-Hilbert action coupled to the canonical scalar field $\phi$ with the potential $V(\phi)$.
	
Since the conformal transformation preserves the causal structure of the spacetime, and $g^E_{ab}$ is asymptotically flat as long as $g_{ab}$ is asymptotically flat, the conformal transformation (\ref{conformal transformation}) of an asymptotically flat black hole spacetime remains an asymptotically flat black hole spacetime with the same event horizon, at least the conformal factor is regular on the horizon\cite{Jacobson:1993pf,Jacobson:1995uq}. Moreover, the conformal factor $\omega = \left[f'(\varphi)\right]^{\frac{1}{D-2}} = \left[f'(R)\right]^{\frac{1}{D-2}}$ is independent of the Killing time $\tau$ in the stationary black hole spacetime, so the conformal transformation (\ref{conformal transformation}) of a stationary black hole solution is still a stationary black hole solution, with the horizon Killing field given by $\xi^a$ up to a normalization factor. If the horizon Killing field $\xi^a$ is normalized as $\xi^a = (\partial_t)^a + \Omega_\mathcal{H} (\partial_\theta)^a$ with $(\partial_t)^a$ representing time translations at infinity in the original frame, we normalize $\xi_E^a = \alpha \xi^a$ with the normalization constant $\alpha$ determined by
\begin{equation}
    \left. g^E_{ab} (\partial_{t^E})^a (\partial_{t^E})^b \right|_{r \to \infty} =
    \left. g^E_{ab} \left[\alpha (\partial_t)^a\right] \left[\alpha (\partial_t)^b\right] \right|_{r \to \infty} = - \left[f'(0)\right]^{\frac{2}{D-2}} \alpha^2 = -1,
\end{equation}
such that $(\partial_{t^E})^a = \alpha (\partial_t)^a = \left[f'(0)\right]^{-\frac{1}{D-2}} (\partial_t)^a $ generates time translations at infinity in the Einstein frame. With the choice $\alpha = \left[f'(0)\right]^{-\frac{1}{D-2}}$, we can investigate the relation of the surface gravity of the black hole in the Einstein frame $\kappa^E$ to that in the original frame $\kappa$. Notice that
\begin{equation}
    \begin{split}
        \nabla^E_a \left(g^E_{bc} \xi_E^b \xi_E^c\right) 
    	&= \nabla_a \left(\omega^2 \alpha^2 g_{bc} \xi^b \xi^c\right)
    	\overset{\mathcal{H}^+}{=} \omega^2 \alpha^2 \nabla_a \left( g_{bc} \xi^b \xi^c\right)  \\
    	&\overset{\mathcal{H}^+}{=}-2 \omega^2 \alpha^2 \kappa g_{ab} \xi^b
    	= -2 \kappa^E g^E_{ab} \xi_E^b.
    \end{split}
\end{equation}
Therefore the relation between $\kappa^E$ and $\kappa$ reads $\kappa^E = \alpha \kappa = \left[f'(0)\right]^{-\frac{1}{D-2}} \kappa$. After the conformal transformation, the null affine parameter $v^E$ on the horizon in the Einstein frame is related to that in the original frame $v$ by\cite{Wald:1984rg}
\begin{equation}
    \frac{\mathrm{d} v^E}{\mathrm{d} v} = c \left[f_R(x,v)\right]^{\frac{2}{D-2}},
\end{equation}
where $c$ is an arbitrary constant and $f_R(x,v)$ is the value of $f'(\varphi) = f'(R)$ at $(x,v)$ on the horizon. As a result,
\begin{equation}\label{relation}
    v^E = c \int_{0}^{v} \mathrm{d} v' \left[f_R(x,v')\right]^{\frac{2}{D-2}}.
\end{equation}
Since $0 = \partial_\tau f'(R) = \kappa v \partial_v f_R(x,v)$ in the stationary background, $f_R(x,v)$ does not depend on $v$, so $v^E$ is proportional to $v$ on the horizon of the stationary black hole. 
	
\subsection{The physical process first law}
	
Since null geodesics are conformally invariant and the Einstein equation holds for $g^E_{ab}$ in the Einstein frame, the derivation of Sec. \ref{sec3} is still tenable for $f(R)$ gravity in the Einstein frame. Let
\begin{equation}
    \mathcal{T}_{ab} = T^E_{ab} + \partial_a \phi \partial_b \phi - g^E_{ab} \left[\frac{1}{2} g_E^{cd} \partial_c \phi \partial_d \phi +V(\phi) \right]
\end{equation}
denote the sum of the stress-energy tensor of the external matter fields in the Einstein frame $T^E_{ab} = - \frac{2}{\sqrt{-g^E}} \frac{\delta S_M}{\delta g_E^{ab}} = T_{ab} / f_R(x,v)$ as well as that of the auxiliary scalar field.
Notice that
\begin{equation}
	\begin{split}
		  \delta \mathcal{T}_{ab} \xi_E^a  k_E^b 
             &\overset{\mathcal{H}^+}{=} \delta T^E_{ab} \xi_E^a  k_E^b  
		  + 2 \partial_a \phi  \left(\partial_b \delta \phi \right) \xi_E^a  k_E^b  \\
		  &- \delta g^E_{ab} \left[ \frac{1}{2} g_E^{cd} \partial_c \phi \partial_d \phi +V(\phi) \right] \xi_E^a  k_E^b  \\
		  &- g^E_{ab} \delta \left[ \frac{1}{2} g_E^{cd} \partial_c \phi \partial_d \phi +V(\phi) \right] \xi_E^a  k_E^b  \\
		  &\overset{\mathcal{H}^+}{=} \delta T^E_{ab} \xi_E^a  k_E^b,
	\end{split}
\end{equation}
where we have used the fact that $k_E^a \partial_a \phi = 0$ and $\left. g^E_{ab} k^a k^b \right|_{\mathcal{H}^+} = 0$ in the stationary background as well as the gauge condition $\delta g_{ab} k_E^a k_E^b \overset{\mathcal{H}^+}{=} 0$ for the perturbations. Thus, only the variation of the external matter fields contributes to the first-order variation of $\mathcal{T}_{ab}$. For perturbations satisfying the gauge conditions listed in Sec. \ref{sec2} that fix $k_E^a$ on the horizon, the non-stationary physical process first law for $f(R)$ gravity in the Einstein frame reads	
\begin{equation}\label{PPFL in Einstein Frame}
    \frac{\kappa^E}{2\pi} \Delta \delta S^E_{\text{dyn}} =  
    \int_{v^E_1}^{v^E_2} \mathrm{d}v^E \int_{\mathcal{C}(v^E)} \mathrm{d} A^E \
    \delta T^E_{ab} \xi_E^a  k_E^b,
\end{equation}
where the dynamical black hole entropy formula for $f(R)$ gravity in the Einstein frame is given by
\begin{equation}\label{DBH entropy in E}
    S^E_{\text{dyn}}[\mathcal{C}]  = \frac{1}{4G_D}
    \left(1 - v^E \frac{\mathrm{d}}{\mathrm{d} v^E}\right)
    \int_{\mathcal{C}(v^E)} \mathrm{d}^{D-2} x \sqrt{\gamma^E}
\end{equation}
with $\gamma^E = \gamma \left[f'(R)\right]^2$ representing the determinant of the induced metric on $\mathcal{C}(v^E)$ in the Einstein frame. 
	
Next we would like to recast the physical process first law and the formula for the dynamical black hole entropy in the original frame. As the perturbation performed on the black hole is non-stationary, the gauge conditions that fix $k_E^a$ will not generally fix $k^a$ in the original frame. Since for non-stationary perturbations, $f_R(x,v)$ is not necessary a constant along the null geodesics on $\mathcal{H}^+$ and $v^E$ presumably would not be proportional to $v$ accordingly. While as mentioned in Sec. \ref{sec2}, the essence of those gauge conditions lies in how to identify spacetime points in two slightly different spacetimes. Those gauge conditions that fix $k_E^a$ under the perturbations imply that we correspond points with the same $v^E$ in the two spacetimes. To rewrite (\ref{PPFL in Einstein Frame}) and (\ref{DBH entropy in E}) in the original frame, in what follows we are going to compare spacetime points sharing the same $v$ on the horizon, and replace $v^E$ with $v$ in the formula of $S^E_{\text{dyn}}$.

Without loss of generality, in the following discussion we set $v^E = v$ in the stationary background, and perform the conformal transformation (\ref{conformal transformation}) on the perturbed geometry $g_{ab} + \delta g_{ab}$ to obtain the non-stationary black hole solution in the Einstein frame. Then $v^E = v + \delta V(x,v)$ on the horizon according to (\ref{relation}) in the perturbed geometry, where $\delta V(x,v)$ is a correction function derived from expanding the conformal factor $\omega^2 = \left[f_R(x,v)\right]^{\frac{2}{D-2}}$ with respect to the non-stationary perturbation of the metric, and it is of first order in perturbation theory. To begin with, let $F(x,v)$ represent a quantity on the horizon, such as the entropy $S_{\text{dyn}}$ or the component of the stress-energy tensor $T_{vv} (x,v)$, which is stationary in the background, and let $\tilde{F}(x,v)$ denote the corresponding quantity in the perturbed spacetime, then $\delta F(x,v) = \tilde{F}(x,v) - F(x,v)$ does not change to first order as we switch from $v^E-$identification to $v-$identification, because
\begin{equation}\label{identification}
    \delta_{v^E} F - \delta_v F = \tilde{F}(x,v - \delta V) - \tilde{F}(x,v) 
    \sim -\partial_v  \tilde{F} \ \delta V = \mathcal{O}(\epsilon^2),
\end{equation}
where $\delta_v$ stands for the comparison between points with the same $v$ in the stationary background and in the perturbed spacetime, and the same connotation holds for $\delta_{v^E}$. We have also used that $\partial_v \tilde{F}$ is of first order in the perturbed spacetime as $F$ is stationary in the unperturbed background. Therefore (\ref{PPFL in Einstein Frame}) does not change to
first order under the reidentification of spacetime points.
In addition, although the difference $\delta V$ between $v$ and $v^E$ leads to first-order corrections to the integral over $v$ and to the null generator $k^a$ in the perturbed geometry, the right-hand side of (\ref{PPFL in Einstein Frame}) will not change to first order if we replace $k^a$ by $k_E^a$ and replace $\int_{v^E_1}^{v^E_2} \mathrm{d} v^E$ by $\int_{v_1}^{v_2} \mathrm{d} v$, since $\delta T^E_{ab}$ is a first-order small quantity and this substitution leads to second-order corrections.
Moreover, in the perturbed geometry,
\begin{equation}
    \begin{split}
		S^E_\text{dyn}[\mathcal{C}(v^E)] 
        &= \frac{1}{4G_D} 
		\left(1 - v^E \frac{\mathrm{d}}{\mathrm{d} v^E}\right)
		\int_{\mathcal{C}} \mathrm{d}^{D-2} x \sqrt{\gamma^E(x,v^E)}  \\
        &= \frac{1}{4G_D} \int_{\mathcal{C}} \mathrm{d}^{D-2} x \left[1 - \left(v+\delta V\right) \left(1-\frac{\partial \delta V}{\partial v^E}\right) \frac{\partial}{\partial v}\right]
        \sqrt{\gamma}f'(R)  \\
        &= \frac{1}{4G_D} 
		\left(1 - v \frac{\mathrm{d}}{\mathrm{d} v}\right)
		\int_{\mathcal{C(v)}} \mathrm{d}^{D-2} x 
        \sqrt{\gamma}f'(R) +  \mathcal{O}(\epsilon^2),
    \end{split}
\end{equation}  
where we have used $\frac{\mathrm{d}}{\mathrm{d}v}\int_{\mathcal{C}} \mathrm{d}^{D-2} x \sqrt{\gamma^E} \sim \mathcal{O}(\epsilon)$ in the third line. Thus replacing $v^E$ with $v$ in (\ref{DBH entropy in E}) leads to second-order corrections. For perturbations left $k^a$ invariant on the horizon, we rewrite (\ref{PPFL in Einstein Frame}) in the original frame as
\begin{equation}\label{PPFL in Original Frame}
    \frac{\kappa}{2\pi} \Delta \delta S_{\text{dyn}} =  
	\int_{v_1}^{v_2} \mathrm{d}v \int_{\mathcal{C}(v)} \mathrm{d} A \
	\delta T_{ab} \xi^a k^b,
\end{equation}
where $S_{\text{dyn}}$ is given by
\begin{equation}
    S_{\text{dyn}}[\mathcal{C}]  = \frac{1}{4G_D}
    \left(1 - v \frac{\mathrm{d}}{\mathrm{d} v}\right)
    \int_{\mathcal{C}(v)} \mathrm{d}^{D-2} x \sqrt{\gamma} f'(R).
\end{equation}
Our results agree with those obtained by means of the Noether charge method\cite{Hollands:2024vbe,Visser:2024pwz}. Moreover, if the stress-energy tensor of the external matter field satisfies the null energy condition $\delta T_{ab} k^a k^b \ge 0$, the dynamical black hole entropy will obey the linearized second law $\Delta \delta S_{\text{dyn}} \ge 0$ under the perturbation.		
	
\subsection{The comparison first law}
	
The conformal transformation (\ref{conformal transformation}) can also be used to derive the non-stationary comparison first law for $f(R)$ gravity. As shown above, the variation of the auxiliary scalar field does not contribute to the first-order variation of $\mathcal{T}_{ab}$. For source-free perturbations of the stationary black hole $\delta T^E_{ab} = 0$, in the Einstein frame we write down the comparison first law for an arbitrary cross-section (\ref{CFL}) as 
\begin{equation}
    \frac{\kappa^E}{2 \pi} \delta S^E_{\text{dyn}}[\mathcal{C}(v^E)] = \delta M^E - \Omega_\mathcal{H} \delta J^E.
\end{equation}
Since the asymptotic forms of $g_{ab}$ and $g^E_{ab}$ agree up to a normalization factor, the mass and angular momenta of the two spacetimes agree up to a normalization factor according to the Iyer-Wald definitions\cite{Jacobson:1995uq,Iyer:1994ys}. And the angular velocities are in agreement as $\xi_E^a \propto \xi^a$. One can determine the relationship between $M^E$ and $M$ as $M^E = \alpha M$ since the time translations at infinity are related as $(\partial_{t^E})^a = \alpha (\partial_t)^a$, and similarly we have $J^E = \alpha J$. Notice that $\kappa^E = \alpha \kappa$, $\delta S_{\text{dyn}}$ does not change to first order as we switch from $v^E-$identification to $v-$identification, and $S^E_{\text{dyn}}$ remains unchanged to first order if we use $v$ instead of $v^E$ in the formula. The comparison first law for an arbitrary cross-section is translated into the claim in the original frame as
\begin{equation}
    \frac{\kappa}{2 \pi} \delta S_{\text{dyn}}[\mathcal{C}(v)] = \delta M - \Omega_\mathcal{H} \delta J.
\end{equation}

\subsection{The second law for vacuum perturbations}
	
For vacuum perturbations we have no external matter fields and $\delta T_{ab} = 0$, so the dynamical black hole entropy is invariant to first order. In order to study the nontrivial leading-order behavior of $S_{\text{dyn}}$, we have to pay attention to the second order in perturbation theory. The first-order variation of the Raychaudhuri equation in the Einstein frame reads
\begin{equation}
    \frac{\mathrm{d}}{\mathrm{d} v^E} \delta \theta^E = 0.
\end{equation}
Because $\theta^E$ vanishes asymptotically at late times, $\delta \theta^E = 0$ holds on $\mathcal{H}^+$. Notice that the stress-energy tensor of the auxiliary scalar field contributes to the second-order variation of $\mathcal{T}_{ab}$. We vary the Raychaudhuri equation combined with the Einstein equation twice to obtain
\begin{equation}
    \frac{\mathrm{d}}{\mathrm{d} v^E} \delta^2 \theta^E
    = -2 \delta \sigma^E_{ab} \delta \sigma_E^{ab}
    - 16\pi G_D \left(k_E^a \partial_a \delta \phi\right)^2.
\end{equation}
And then integrating this equation on $\mathcal{H}^+$ between $\mathcal{C}(v^E_1)$ and $\mathcal{C}(v^E_2)$ returns
\begin{equation}\label{2nd-order behavior in E}
    \frac{\kappa^E}{2\pi} \Delta \delta^2 S^E_{\text{dyn}} =  \frac{1}{4 \pi G_D} \int_{v^E_1}^{v^E_2} \mathrm{d}v^E \int_{\mathcal{C}(v^E)} \mathrm{d} A^E
    \kappa^E v^E \left[\delta \sigma^E_{ab}\delta \sigma_E^{ab} + 8 \pi G_D \left(k_E^a \partial_a \delta \phi\right)^2\right].
\end{equation}
Since $\delta \sigma^E_{ab}\delta \sigma_E^{ab} \ge 0$ and $\left(k_E^a \partial_a \delta \phi\right)^2 \ge 0$, for vacuum perturbations, the dynamical black hole entropy satisfies the second law at the second order in the Einstein frame. In what follows we would like to explore the second-order behavior of the dynamical black hole entropy for vacuum perturbations in the original frame based on this equation.
	
First, (\ref{2nd-order behavior in E}) is invariant to second order in the perturbation theory as we decide to compare spacetime points sharing the same $v$ rather than $v^E$ on the horizon. Since by the argument similar to that of (\ref{identification}), switching from $v^E-$identification to $v-$identification for $\delta S_{\text{dyn}}$ and $\delta^2 S_{\text{dyn}}$ leads to third-order corrections as $\partial_v S_{\text{dyn}} \sim \mathcal{O}(\epsilon^2)$, and altering the identification of spacetime points also leads to third-order corrections on the right-hand side of (\ref{2nd-order behavior in E}). Moreover, for vacuum perturbations $S^E_{\text{dyn}}$ is invariant to second order if we replace $v^E$ with $v$ in the formula. As $S^E_{\text{dyn}}$ is a constant to first order under the vacuum perturbations, in the perturbed spacetime
\begin{equation}
    \left(1-v^E \frac{\mathrm{d}}{\mathrm{d} v^E}\right) S^E_{\text{BH}} =  C + \mathcal{O}(\epsilon^2),
\end{equation}
where $S^E_{\text{BH}} = \frac{1}{4 G_D} \int_{\mathcal{C}} \mathrm{d}^{D-2} x \sqrt{\gamma^E}$ is the Bekenstein-Hawking entropy in the Einstein frame and $C$ is a constant. The solution to this equation reads
\begin{equation}
    S^E_{\text{BH}} = C + \varepsilon v^E + \mathcal{O}(\epsilon^2),
\end{equation}
where $\varepsilon$ is a constant. Requiring that $S^E_{\text{BH}}$ remains finite at future infinity $v^E \to \infty$ in the perturbed spacetime leads to the conclusion that 
\begin{equation}
    S^E_{\text{BH}} = C + \mathcal{O}(\epsilon^2)
\end{equation}
Thus $\frac{\mathrm{d}}{\mathrm{d} v^E} S^E_{\text{BH}}$ is of second order for vacuum perturbations, and replacing $v^E$ with $v$ in the expression of $S^E_{\text{dyn}}$ leads to third-order corrections.
	
Next, we wish to rewrite (\ref{2nd-order behavior in E}) in the original frame. Notice that $\delta \theta^E \overset{\mathcal{H}^+}{=} 0$ for vacuum perturbations in the Einstein frame. In the original frame we have
\begin{equation}
    \begin{split}
		\delta \partial_v \ln \left[\sqrt{\gamma} f'(R)\right]
		&= \frac{f'(R) \partial_v \delta \sqrt{\gamma} + \sqrt{\gamma} f''(R) \partial_v \delta R}{\sqrt{\gamma} f'(R)} \\
		&= \delta \theta
		+ \frac{f''(R)}{f'(R)} \partial_v \delta R \overset{\mathcal{H}^+}{=} 0.
    \end{split}
\end{equation}		
As a result, to first order $\delta \sigma^E_{ab}$ in the original frame reads
\begin{equation}
    \begin{split}
		\delta \sigma^E_{ab} &= \frac{1}{2} \mathcal{L}_{k^E} \delta \gamma^E_{ab} - \frac{1}{D-2} \delta \theta^E \gamma_{ab}  \\
		&= \frac{1}{2} \left[f'(\varphi)\right]^{\frac{2}{D-2}} \left[\mathcal{L}_{k} \delta \gamma_{ab} + \frac{2}{D-2} \frac{f''(\varphi)}{f'(\varphi)} \gamma_{ab} \mathcal{L}_{k} \delta \varphi\right]  \\
		&= \left[f'(\varphi)\right]^{\frac{2}{D-2}}
		\left[\frac{1}{2} \mathcal{L}_{k} \delta \gamma_{ab} - \frac{1}{D-2} \delta \theta \gamma_{ab} \right]  \\
		&= \left[f'(\varphi)\right]^{\frac{2}{D-2}} \delta \sigma_{ab},
    \end{split}
\end{equation}
where the second equality follows from $\delta \theta^E \overset{\mathcal{H}^+}{=} 0$ as well as $\gamma^E_{ab} = \left[f'(\varphi)\right]^{\frac{2}{D-2}} \gamma_{ab}$, and the third follows from $\delta \theta
\overset{\mathcal{H}^+}{=} - \frac{f''(R)}{f'(R)} \partial_v \delta R = - \frac{f''(\varphi)}{f'(\varphi)} \mathcal{L}_{k} \delta \varphi $. We also emphasize that $\mathcal{L}_{k}$ only acts on $\delta \gamma_{ab}$ and $\delta \varphi$ as they are non-stationary. Thus
\begin{equation}
    \begin{split}
        \delta \sigma^E_{ab} \delta \sigma_E^{ab}
        &= \gamma_E^{ac} \gamma_E^{bd} \delta \sigma^E_{ab} \delta \sigma^E_{cd}
		= \gamma_E^{ac} \gamma_E^{bd} \left[f'(\varphi)\right]^{\frac{4}{D-2}}
		\delta \sigma_{ab}\delta \sigma_{cd} \\
        &= \gamma^{ac} \gamma^{bd} \delta \sigma_{ab} \delta \sigma_{cd}
		= \delta \sigma_{ab}\delta \sigma^{ab}.
    \end{split}
\end{equation}	
Furthermore, 
\begin{equation}
    \delta \phi =\frac{1}{\sqrt{16 \pi G_D}} \sqrt{\frac{2 (D-1)}{D-2}} \delta \ln f'(\varphi) 
	= \frac{1}{\sqrt{16 \pi G_D}} \sqrt{\frac{2(D-1)}{D-2}} \frac{f''(R)}{f'(R)} \delta R,
\end{equation}
so $(k_E^a \partial_a \phi)^2$ in the original frame reads
\begin{equation}
    (k_E^a \partial_a \phi)^2 = \frac{D-1}{8\pi G_D(D-2)} \left(\frac{f''(R)}{f'(R)}\right)^2 \left(\partial_v \delta R\right)^2 
	= \frac{D-1}{8\pi G_D(D-2)} \left(\delta \theta\right)^2.
\end{equation}
Gathering all of the results obtained above, we rewrite (\ref{2nd-order behavior in E}) in the original frame as
\begin{equation}\label{MCEF for FR}
    \frac{\kappa}{2 \pi} \Delta \delta^2 S_{\text{dyn}} 
    = \frac{1}{4 \pi G_D} \int_{v_1}^{v_2} \mathrm{d} v \int_{\mathcal{C}(v)} \mathrm{d} A \
    \kappa v f'(R) \left[\delta \sigma_{ab}\delta \sigma^{ab} + \frac{D-1}{D-2} \left(\delta \theta\right)^2 \right].
\end{equation}
This result aligns with the outcome obtained by calculating the modified canonical energy flux for $f(R)$ gravity, which is shown in Appendix \ref{appendix}. As $f'(R) > 0$, $\delta \sigma_{ab}\delta \sigma^{ab} \ge 0$ and $(\delta \theta)^2 \ge 0$, we may also verify the second law for vacuum perturbations at second order in the original frame.

\subsection{Relation to Wald entropy of the generalized apparent horizon}
	
If the derivative of the Wall entropy with respect to the affine parameter $\mathrm{d} S_{\text{Wall}} / \mathrm{d}v$ is positive, which follows from the second law of black hole thermodynamics\cite{Jacobson:1995uq,Wall:2015raa}, it suggests that dynamical black hole entropy is associated to the entropy of a surface inside the black hole. Although the location of the apparent horizon associated with the prescribed cross-section $\mathcal{C}(v)$ is ambiguous as it depends on the choice of the simultaneous surface, for perturbations of a stationary black hole with bifurcate Killing horizon, the notion of the area of an apparent horizon corresponding to cross-section $\mathcal{C}(v)$ is well defined to first order\cite{Hollands:2024vbe}. And it was shown in the appendix A of \cite{Hollands:2024vbe} that the dynamical black hole entropy is equal to the area entropy of the apparent horizon to first order in GR. Visser and Yan also provided a more pedagogical proof of this claim using Gaussian null coordinates (GNC) system\cite{Visser:2024pwz}. Their proof can be applied to $f(R)$ gravity in the Einstein frame since it's only a geometric fact. And in the Einstein frame this relation reads
\begin{equation}\label{Relation in Einstein frame}
    S^E_{\text{dyn}} = \frac{1}{4G_D} \left(1 - v^E \frac{\mathrm{d}}{\mathrm{d} v^E}\right) A\left[\mathcal{C}(v^E)\right] 
    = \frac{A\left[\mathcal{T}^E(v^E)\right]}{4G_D},
\end{equation}
where $\mathcal{T}^E(v^E)$ is a constant$-v^E$ surface of the apparent horizon in the Einstein frame $\mathcal{A}^E$, which is determined by the condition that the outgoing null expansion vanishes
\begin{equation}\label{condition}
    \theta^E_{\tilde{k}} = \frac{\partial}{\partial {\lambda^E}} \ln \sqrt{\gamma^E} = 0,
\end{equation}  
where $\lambda^E$ is the affine parameter of the (future directed) outgoing null normal $\tilde{k}_E^a$ to $\mathcal{T}^E(v^E)$, $A\left[\mathcal{C}(v^E)\right]$ is the area of the horizon cross-section in the Einstein frame, and $A\left[{\mathcal{T}^E}(v^E)\right]$ is the area of $\mathcal{T}^E(v^E)$. We wish to recast this relation in the original frame. As conformal transformations preserve orthogonality as well as the null property, the condition that locates the apparent horizon $\mathcal{T}^E(v^E)$ in the Einstein frame (\ref{condition}) is translated into
\begin{equation}
    \frac{1}{c \left[f'(R)\right]^{\frac{2}{D-2}}}
    \frac{\partial}{\partial \lambda} \ln \left[\sqrt{\gamma} f'(R) \right] = 0
\end{equation} 
in the original frame, where $c$ is a constant and $\lambda$ is the affine parameter of the outgoing null normal in the original frame. The \emph{generalized expansion} for $f(R)$ gravity in the original frame is defined as\cite{Matsuda:2020yvl}
\begin{equation}
    \Theta_{\lambda} = \frac{\partial}{\partial \lambda} \ln \left[\sqrt{\gamma} f'(R) \right].
\end{equation}
Then the constant$-v^E$ surface of the apparent horizon in the Einstein frame $\mathcal{T}^E(v^E)$ is translated into the constant$-v$ surface of the \emph{generalized apparent horizon}  $\mathcal{T}(v)$ in the original frame, which is defined as the $D-2$ dimensional section with vanishing outgoing generalized expansion. As we have shown above, $S_{\text{dyn}}$ is invariant to first order if we replace $v^E$ with $v$ in its formula. Therefore, the relation (\ref{Relation in Einstein frame}) in the original frame is rewritten as
\begin{equation}
    S_{\text{dyn}} = \frac{1}{4G_D} \left(1 - v \frac{\mathrm{d}}{\mathrm{d} v}\right) \int_{\mathcal{C}(v)} \mathrm{d}^{D-2} x \sqrt{\gamma} f'(R) 
    = \frac{1}{4G_D} \int_{\mathcal{T}(v)} \mathrm{d}^{D-2} x \sqrt{\gamma} f'(R).
\end{equation}
Thus the dynamical black hole entropy is equal to the Wald entropy of the generalized apparent horizon in the original frame of $f(R)$ gravity.

    
\section{Dynamical black hole entropy in canonical scalar-tensor theory}\label{sec5}
	
In this section we apply the methodology of the previous section to the canonical scalar-tensor theory. Firstly we derive both the physical process first law and the comparison first law for the canonical scalar-tensor theory. Secondly we study the second-order behavior of the dynamical black hole entropy for vacuum perturbations in the Einstein frame, and then convert it to the expression in the original frame. Finally we determine the relationship between the dynamical black hole entropy of the cross-section and the Wald entropy of the generalized apparent horizon in the original frame.
	
\subsection{The conformal transformation and the Einstein frame}
The canonical scalar-tensor theory is described by the action
\begin{equation}
    I_{\text{st}} = \frac{1}{16 \pi G_D} \int \mathrm{d}^D x \sqrt{-g}
    \left[F(\varphi) R - \frac{1}{2} g^{ab} \partial_a \varphi \partial_b \varphi - U(\varphi) \right].
\end{equation}
And we assume that $F(\varphi) > 0$ to avoid tensor ghosts\cite{Matsuda:2020yvl}. We turn to the Einstein frame and redefine the scalar field as\cite{Dicke:1961gz}
\begin{equation}\label{transformation for scalar-tensor theory}
    \begin{split}
        g^E_{ab} &\equiv \left[F(\varphi)\right]^{\frac{2}{D-2}} g_{ab}, \\ 
		\phi &\equiv \frac{1}{\sqrt{16 \pi G_D}} \int \mathrm{d} \tilde{\varphi} \
		\sqrt{F(\tilde{\varphi})^{-1} + \frac{2(D-1)}{D-2} \left[\frac{F'(\tilde{\varphi})}{F(\tilde{\varphi})}\right]^2 }.
    \end{split}
\end{equation}
Then the action can be rewritten as
\begin{equation}
    I_{\text{st}} =  \int \mathrm{d}^D x \sqrt{-g^E}
    \left[ \frac{R^E}{16 \pi G_D} - \frac{1}{2} g_E^{ab} \partial_a \phi \partial_b \phi - V(\phi) \right],
\end{equation}
where
\begin{equation}
    V(\phi) = \frac{1}{16 \pi G_D} \left[F(\varphi)\right]^{-\frac{D}{D-2}} U(\varphi).
\end{equation}   
And for an asymptotically flat stationary black hole solution to the scalar-tensor theory, the spacetime given by (\ref{transformation for scalar-tensor theory}) remains an asymptotically flat stationary black hole with the same event horizon. Without loss of generality we assume $\varphi$ tends to be a constant at spatial infinity. So the surface gravities of the two black hole spacetimes are related as $\kappa^E = \alpha \kappa$ with $\alpha$ representing the value of $\left[F(\varphi)\right]^{-\frac{1}{D-2}}$ at infinity.
	
\subsection{The physical process first law}
The derivation of the physical process first law for scalar-tensor theory is  similar to the previous situation for $f(R)$ gravity. First, for the perturbations left $k_E^a$ invariant, we write down the non-stationary physical process first law in the Einstein frame as
\begin{equation}\label{PPFL in Einstein Frame for Scalar-tensor Theory}
    \frac{\kappa^E}{2\pi} \Delta \delta S^E_{\text{dyn}} =  
    \int_{v^E_1}^{v^E_2} \mathrm{d}v^E \int_{\mathcal{C}(v^E)} \mathrm{d} A^E \
    \delta T^E_{ab} \xi_E^a  k_E^b,
\end{equation}
where
\begin{equation}
    S^E_{\text{dyn}}[\mathcal{C}]  = \frac{1}{4G_D}
    \left(1 - v^E \frac{\mathrm{d}}{\mathrm{d} v^E}\right)
    \int_{\mathcal{C}(v^E)} \mathrm{d}^{D-2} x \sqrt{\gamma^E}.
\end{equation}
Then we would like to recast this equation in the original frame. By the arguments similar to the last section, switching from $v^E-$identification to $v-$identification leads to second-order corrections. And for perturbations left $k^a$ invariant, we recast (\ref{PPFL in Einstein Frame for Scalar-tensor Theory}) in the original frame as
\begin{equation}
    \frac{\kappa}{2\pi} \Delta \delta S_{\text{dyn}} =  
    \int_{v_1}^{v_2} \mathrm{d}v \int_{\mathcal{C}(v)} \mathrm{d} A \
    \delta T_{ab} \xi^a k^b,
\end{equation}
where
\begin{equation}
    S_{\text{dyn}}[\mathcal{C}]  = \frac{1}{4G_D}
    \left(1 - v \frac{\mathrm{d}}{\mathrm{d} v}\right)
    \int_{\mathcal{C}(v)} \mathrm{d}^{D-2} x \sqrt{\gamma} F(\varphi).
\end{equation}
	
\subsection{The comparison first law}
	
Next, the derivation of the comparison first law for scalar-tensor theory is straightforward. For source-free perturbations $\delta T^E_{ab}  = 0$ we write the comparison first law (\ref{CFL}) in the Einstein frame as
\begin{equation}
    \frac{\kappa^E}{2 \pi} \delta S^E_{\text{dyn}}[\mathcal{C}(v^E)] = \delta M^E - \Omega_\mathcal{H} \delta J^E.
\end{equation}
Then as $\kappa^E = \alpha \kappa$, $M^E = \alpha M$ and $J^E = \alpha J$ with $\alpha$ representing the inverse of the conformal factor at spatial infinity, we obtain the comparison first law in the original frame
\begin{equation}
    \frac{\kappa}{2 \pi} \delta S_{\text{dyn}}[\mathcal{C}(v)] = \delta M - \Omega_\mathcal{H} \delta J.
\end{equation}
	
\subsection{The second law for vacuum perturbations}
	
If external matter fields are absent, for vacuum perturbations $\delta T^E_{ab} = 0$ it follows immediately that the dynamical black hole entropy is a constant to first order. So we have to pay attention to the second-order behavior to study the nontrivial change of the dynamical black hole entropy. The first-order variation of the Raychaudhuri equation implies that $\delta \theta^E$ is 0 along the horizon. We consider the second-order variation of the Raychaudhuri equation 
\begin{equation}
    \frac{\mathrm{d}}{\mathrm{d} v^E} \delta^2 \theta^E
    = -2 \delta \sigma^E_{ab}\delta \sigma_E^{ab}
    - 16\pi G_D \left(k_E^a \partial_a \delta \phi\right)^2,
\end{equation}
and then integrate it on the horizon between two arbitrary cross-sections to obtain
\begin{equation}\label{2nd-order in E for ST}
    \frac{\kappa^E}{2\pi} \Delta \delta^2 S^E_{\text{dyn}} =  \frac{1}{4 \pi G_D} \int_{v^E_1}^{v^E_2} \mathrm{d}v^E \int_{\mathcal{C}(v^E)} \mathrm{d} A^E \
    \kappa^E v^E 
    \left[\delta \sigma^E_{ab}\delta \sigma_E^{ab} + 8 \pi G_D \left(k_E^a \partial_a \delta \phi\right)^2\right].
\end{equation}
As $\delta \sigma^E_{ab}\delta \sigma_E^{ab} \ge 0$ and $\left(k_E^a \partial_a \delta \phi\right)^2 \ge 0$, the second law holds at second order for vacuum perturbations in the Einstein frame. Next we are going to rewrite this equation in the original frame.
	
By the arguments similar to those in the previous section, (\ref{2nd-order in E for ST}) is invariant to second order as we recast it in the original frame and switch from $v^E-$identification to $v-$identification on the horizon. Notice that $\delta \theta^E \overset{\mathcal{H}^+}{=} 0$ in the Einstein frame. In the original frame we obtain
\begin{equation}
    \delta \partial_v \ln \left[\sqrt{\gamma}F(\varphi)\right] 
    = \delta \theta + \frac{F'(\varphi)}{F(\varphi)} \partial_v \delta \varphi \overset{\mathcal{H}^+}{=} 0.
\end{equation}
As a result,
\begin{equation}
    \begin{split}
		\delta \sigma^E_{ab} &= \frac{1}{2} \mathcal{L}_{k^E} \delta \gamma^E_{ab}  \\
		&= \left[F(\varphi)\right]^{\frac{2}{D-2}} \left[\frac{1}{2}\mathcal{L}_{k} \delta \gamma_{ab} + \frac{1}{D-2} \gamma_{ab} \frac{F'(\varphi)}{F(\varphi)} \mathcal{L}_k \delta \varphi\right] \\
		&= \left[F(\varphi)\right]^{\frac{2}{D-2}} \left[\frac{1}{2}\mathcal{L}_{k} \delta \gamma_{ab} - \frac{1}{D-2} \delta \theta \gamma_{ab} \right]  \\
		&= \left[F(\varphi)\right]^{\frac{2}{D-2}} \delta \sigma_{ab}.
    \end{split}
\end{equation}
Furthermore, 
\begin{equation}
    k_E^a \partial_a \delta \phi = \frac{\mathrm{d} \phi}{\mathrm{d} \varphi} \partial_v \delta \varphi 
    = \frac{1}{\sqrt{16 \pi G_D}} \sqrt{F(\varphi)^{-1} + \frac{2(D-1)}{D-2}\left[\frac{F'(\varphi)}{F(\varphi)}\right]^2} \partial_v \delta \varphi.
\end{equation}
Thus (\ref{2nd-order in E for ST}) in the original frame reads
\begin{equation}
    \begin{split}
        \frac{\kappa}{2 \pi} \Delta \delta^2 S_{\text{dyn}} &= \frac{1}{4 \pi G_D} \int_{v_1}^{v_2} \mathrm{d}v \int_{\mathcal{C}(v)} \mathrm{d}A \ \kappa v F(\varphi)  \\
		&\times\left[\delta \sigma_{ab}\delta \sigma^{ab} + \frac{D-1}{D-2} \left(\delta \theta\right)^2 + \frac{1}{2 F(\varphi)} \left(k^a\partial_a\delta\varphi\right)^2\right].
    \end{split}
\end{equation}
Since $F(\varphi) > 0$, $\delta \sigma_{ab}\delta \sigma^{ab}$, $\left(\delta \theta\right)^2$ and $\left(k^a\partial_a\delta\varphi\right)^2$ are nonnegative, the second law holds at second order for vacuum perturbations in the original frame.
	
\subsection{Relation to Wald entropy of the generalized apparent horizon}
	
Finally we would like to determine the relation between the dynamical black hole entropy of $\mathcal{C}(v)$ and the entropy of the generalized apparent horizon inside the black hole. The starting point is the relation to the Bekenstein-Hawking entropy of the apparent horizon in the Einstein frame
\begin{equation}\label{Relation in Einstein frame for scalar-tensor theory}
    S^E_{\text{dyn}} = \frac{1}{4G_D} \left(1 - v^E \frac{\mathrm{d}}{\mathrm{d} v^E}\right) A\left[\mathcal{C}(v^E)\right] 
    = \frac{A\left[\mathcal{T}^E(v^E)\right]}{4G_D}.
\end{equation}
We would like to recast this relation in the original frame. Define the generalized expansion for canonical scalar-tensor theory as\cite{Matsuda:2020yvl}
\begin{equation}
    \Theta_{\lambda} = \frac{\partial}{\partial \lambda} \ln \left[\sqrt{\gamma} F(\varphi)\right],
\end{equation}
where $\lambda$ is the affine parameter of the null geodesics in the original frame. We define the generalized apparent horizon for the black hole spacetime of the canonical scalar-tensor theory as the section $\mathcal{T}$ with vanishing outgoing generalized expansion. Relation (\ref{Relation in Einstein frame for scalar-tensor theory}) is translated into
\begin{equation}
    S_{\text{dyn}} = \frac{1}{4G_D} \left(1 - v \frac{\mathrm{d}}{\mathrm{d} v}\right) \int_{\mathcal{C}(v)} \mathrm{d}^{D-2} x \sqrt{\gamma} F(\varphi) 
    = \frac{1}{4G_D} \int_{\mathcal{T}(v)} \mathrm{d}^{D-2} x \sqrt{\gamma} F(\varphi).
\end{equation}
Thus in the original frame the dynamical black hole entropy is equal to the Wald entropy of the generalized apparent horizon for canonical scalar-tensor theory.
	

\section{Conclusion and discussion}\label{sec6}
	
In this paper we first generalize the pedagogical proof of the non-stationary physical process first law to non-vacuum black holes, and give a pedagogical deduction of the second-order behavior for $S_{\text{dyn}}$ under the vacuum perturbations in GR. We then derive both the non-stationary physical process first law and the non-stationary comparison first law for $f(R)$ gravity as well as canonical scalar-tensor theory by means of conformal transformations. Our formulas for dynamical black hole entropy agree with those derived by the Noether charge method. We further study the second-order behavior of $S_{\text{dyn}}$ for vacuum perturbations in those theories, and find that second law is obeyed both in the Einstein frames and in the original frames. Moreover, we determine the relationship between the dynamical black hole entropy and the Wald entropy of the generalized apparent horizon in the original frame.

In the future work we plan to generalize the study of dynamical black hole entropy to more general theories of gravity, such as $f(\text{Ricci})$, based on other field redefinition techniques\cite{Jacobson:1993vj,Dong:2013qoa,MohammadiMozaffar:2016vpf}. So far our arguments for dynamical black hole entropy only dwell on the classical level. It would be important to investigate at the semi-classical level whether the entropy of the Hawking radiation of a black hole will obtain a dynamical correction term after the non-stationary perturbation.


\acknowledgments
This work is partly supported by the National Key Research and Development Program of China with Grant No. 2021YFC2203001 as well as the National Natural Science Foundation of China with Grants No. 12075026, No. 12035016, No. 12361141825, and No. 12375058.

 
\appendix
	
\section{The modified canonical energy flux for \texorpdfstring{$f(R)$}{f(R)} gravity}\label{appendix}
	
In this appendix we calculate the modified canonical energy flux for $f(R)$ gravity using the covariant phase space formalism. To simplify the calculations, we shall follow the gauge conditions of \cite{Hollands:2024vbe} that take $\xi^a$ to be fixed under the variation. And the gauge conditions on $\delta g_{ab}$ at the event horizon reads:
\begin{equation}
    \xi^a \delta g_{ab} \overset{\mathcal{H}^+}{=} 0, \quad
    \nabla_a (\xi^b \xi^c \delta g_{bc}) \overset{\mathcal{H}^+}{=} 0,
\end{equation}
which are less relaxed than the gauge conditions in Sec. \ref{sec2}.
	
\subsection{The covariant phase space formalism and the fundamental identity}
	
First let us briefly review the covariant phase space formalism. Consider an arbitrary diffeomorphism covariant theory of gravity in $n$-dimensions derived from a Lagrangian $n$-form $\boldsymbol{L}$, then the variation of the Lagrangian can always be expressed as
\begin{equation}\label{variation of Lagrangian}
    \delta \boldsymbol{L} = \boldsymbol{E} \delta \phi + \mathrm{d} \delta \boldsymbol{\theta},
\end{equation}
where $\phi$ is the collection of dynamical fields such as metric $g_{ab}$ and other matter fields, $\boldsymbol{E}$ is the equation of motion locally constructed out of $\phi$, and the symplectic potential $(n-1)$-form $\boldsymbol{\theta}(\phi,\delta\phi)$ is locally constructed out of $\phi$, $\delta\phi$ and their derivatives and is linear in $\delta\phi$. The symplectic current $(n-1)$-form is obtained from $\boldsymbol{\theta}$ via\footnote{Here we assume that the field variations $\delta_1 \phi$ and $\delta_2 \phi$ arise from a two-parameter variation $\phi(\lambda_1,\lambda_2)$ and thus commute.}
\begin{equation}\label{symplectic current}
    \boldsymbol{\omega} (\phi; \delta_1 \phi, \delta_2 \phi)=\delta_1 \boldsymbol{\theta}(\phi,\delta_2 \phi)-\delta_2 \boldsymbol{\theta}(\phi,\delta_1 \phi).
\end{equation}	
Let $\chi^a$ be an arbitrary vector field which is also the infinitesimal generator of a diffeomorphism, then the associated Noether current $(n-1)$-form $\boldsymbol{J}$ is defined by
\begin{equation}\label{Noether current}
    \boldsymbol{J} (\phi) = \boldsymbol{\theta} (\phi, \mathcal{L}_\chi \phi) - \chi \cdot \boldsymbol{L} (\phi),
\end{equation}
where the notation $\cdot$ denotes the contraction of a vector field with the first index of a differential form. It was shown that the Noether current can also be written in the form\cite{Iyer:1995kg,Seifert:2006kv}
\begin{equation}\label{Noether charge}
    \boldsymbol{J} = \mathrm{d}\boldsymbol{Q}[\chi] + \chi^a \boldsymbol{C}_a.
\end{equation}
Here, the $(n-2)$-form $\boldsymbol{Q}$ is referred to as the “Noether charge”\cite{Iyer:1994ys} and the dual vector valued $(n-1)$-form $\boldsymbol{C}_a$ vanishes when the equations of motion are satisfied.
Taking the first variation of (\ref{Noether current}) (taking the vector field $\chi^a$ to be fixed) and using (\ref{variation of Lagrangian}) and (\ref{symplectic current}), we obtain
\begin{equation}
    \delta \boldsymbol{J}(\phi) = -\chi \cdot \left[\boldsymbol{E}(\phi) \delta \phi\right] + \boldsymbol{\omega} (\phi; \delta\phi, \mathcal{L}_\chi \phi)+\mathrm{d}\left[\chi \cdot \boldsymbol{\theta}(\phi,\delta \phi)\right].
\end{equation}
Considering the variation of (\ref{Noether charge}), we obtain the fundamental identity according to\cite{Hollands:2012sf}
\begin{equation}\label{fundamental identity}
    \boldsymbol{\omega}(\phi; \delta\phi, \mathcal{L}_\chi \phi) = \chi \cdot \left[\boldsymbol{E}(\phi) \delta \phi\right] +\chi^a \delta \boldsymbol{C}_a (\phi) 
    \mathrm{d}\left[\delta\boldsymbol{Q}[\chi]-\chi \cdot \boldsymbol{\theta}(\phi,\delta \phi)\right].
\end{equation}
For the case where $\chi^a$ is a Killing field of the background $\phi$ such that $\mathcal{L}_\chi \phi = 0$, we vary the fundamental identity (\ref{fundamental identity}) to yield\footnote{Note that here the symbol $\mathcal{L}_\chi \delta$ is seen as “one” variation, which means that
\begin{equation}
    \boldsymbol{\omega} (\phi; \delta \phi, \mathcal{L}_\chi \delta \phi)=\delta \boldsymbol{\theta}(\phi,\mathcal{L}_\chi \delta \phi)-(\mathcal{L}_\chi \delta) \boldsymbol{\theta}(\phi,\delta \phi).
\end{equation}
For instance, we write $\boldsymbol{\theta}(\phi, \delta\phi)=\boldsymbol{D}\delta\phi$ formally, where $\boldsymbol{D}$ is a linear operator valued $(n-1)$-form acting on $\delta\phi$ constructed out of $\phi$, its derivatives, and the covariant derivative with respect to the stationary metric $g_{ab}$. If we view $\mathcal{L}_\chi \delta$ as “one” variation, we will get
\begin{equation}
    (\mathcal{L}_\chi \delta)\boldsymbol{\theta}(\phi, \delta\phi) = (\mathcal{L}_\chi \delta \boldsymbol{D})\delta\phi+\boldsymbol{D}(\mathcal{L}_\chi \delta^2\phi),
\end{equation}
and if we see $\mathcal{L}_\chi \delta$ as two subsequent variations, we will get
\begin{equation}
    \mathcal{L}_\chi (\delta\boldsymbol{\theta}(\phi, \delta\phi)) = (\mathcal{L}_\chi \delta \boldsymbol{D})\delta\phi+\delta \boldsymbol{D}(\mathcal{L}_\chi\delta\phi)+\boldsymbol{D}(\mathcal{L}_\chi \delta^2\phi), 
\end{equation}
where $\delta \boldsymbol{D}$ is constructed out of $\phi$, $\delta\phi$, their derivatives, and the variation of the Christoffel symbols.}
\begin{equation}\label{second variation of fundamental identity}
    \begin{split}
        \boldsymbol{\omega}(\phi; \delta\phi, \mathcal{L}_\chi \delta \phi) &= \chi \cdot \left[\delta\boldsymbol{E}(\phi) \delta \phi\right] + \chi \cdot \left[\boldsymbol{E}(\phi) \delta^2 \phi\right] \\
        &+\chi^a \delta^2 \boldsymbol{C}_a (\phi)+ \mathrm{d}\left[\delta^2\boldsymbol{Q}[\chi]-\chi \cdot \delta \boldsymbol{\theta}(\phi,\delta \phi)\right].
    \end{split}
\end{equation}
	
\subsection{The second-order behavior of \texorpdfstring{$S_{\text{dyn}}$}{S} for vacuum perturbations}
	
Now let's consider the pure gravity theory in which matter fields are absent in the Lagrangian. For vacuum perturbations, we have no external matter sources, $\delta T_{ab}=0$, and (\ref{PPFL in Original Frame}) states that there is no change of dynamical black hole entropy with respect to affine time at first order, $\Delta \delta S_{\text{dyn}}=0$. Thus we must go to second order in perturbation theory to obtain the leading-order dynamical behavior of the black hold entropy. By (\ref{second variation of fundamental identity}) and setting $\chi^a=\xi^a$, in the case that the vacuum equations of motion hold, we obtain\cite{Hollands:2024vbe}
\begin{equation}\label{second variation}
    \begin{split}
         &\boldsymbol{\omega} (g;\delta g, \mathcal{L}_\xi\delta g)+\mathrm{d}\left[\xi \cdot \delta \boldsymbol{\theta}(g,\delta g)-\xi \cdot \delta^2\boldsymbol{B}_\mathcal{H}(g)\right] \\
        &=\mathrm{d}\left[\delta^2\boldsymbol{Q}[\xi] - \xi \cdot \delta^2\boldsymbol{B}_\mathcal{H}(g)\right]=\frac{\kappa}{2\pi}\mathrm{d}\delta^2\boldsymbol{S}_{\text{dyn}}(g),
    \end{split}
\end{equation}
where $\boldsymbol{B}_\mathcal{H}$ satisfying $\boldsymbol{\theta} \overset{\mathcal{H}^+}{=} \delta\boldsymbol{B}_\mathcal{H}$ is defined on $\mathcal{H}$ according to the theorem 1 of \cite{Hollands:2024vbe}, and it has the following form
\begin{equation}\label{definition of B}	 
    \boldsymbol{B}_\mathcal{H}=
    \boldsymbol{\epsilon}^{(n-1)}\sum_{i=0}^m \tilde{T}_{(i)}^{b_1 \cdots b_i cd}\nabla_{(b_1}\cdots\nabla_{b_i)}\mathcal{L}_\xi g_{cd},
\end{equation}
where the tensors $\tilde{T}_{(i)}^{b_1 \cdots b_i cd}=\tilde{T}_{(i)}^{(b_1 \cdots b_i) (cd)}$ are smooth on $\mathcal{H}$ and are locally and covariantly constructed from the metric, curvature, covariant derivatives of the curvature as well as $\xi^a$ and $N^a$, with $\xi^a$ and $N^a$ appearing only algebraically\footnote{The vector field $N^a$ is introduced by\cite{Hollands:2024vbe}
\begin{equation}
    \nabla_a\xi_b\overset{\mathcal{H}^+}{=}2\kappa N_{[a}\xi_{b]}, \quad N^a N_a\overset{\mathcal{H}^+}{=}0, \quad N^a \xi_a\overset{\mathcal{H}^+}{=}1.
\end{equation}}. 
And the entropy $(n-2)$-form $\boldsymbol{S}_{\text{dyn}}$ is defined on $\mathcal{H}$ by
\begin{equation}
    \boldsymbol{S}_{\text{dyn}}\equiv \frac{2\pi}{\kappa}(\boldsymbol{Q}[\xi]-\xi\cdot\boldsymbol{B}_\mathcal{H}). 
\end{equation}
Integrating (\ref{second variation}) between two cross-sections on the horizon, we find that
\begin{equation}\label{second law in second order}
    \frac{\kappa}{2\pi}\Delta\delta^2 S_{\text{dyn}} = \int_{\mathcal{H}_{12}}\boldsymbol{e}_G(g; \delta g, \delta g),
\end{equation}
where the modified canonical energy flux $(n-1)$-form $\boldsymbol{e}_G$ is defined by
\begin{equation}
    \boldsymbol{e}_G(g; \delta g, \delta g) \equiv \boldsymbol{\omega} (g;\delta g, \mathcal{L}_\xi\delta g)
    +\mathrm{d}\left[\xi \cdot \delta \boldsymbol{\theta}(g,\delta g)-\xi \cdot \delta^2\boldsymbol{B}_\mathcal{H}(g)\right].
\end{equation}
From (\ref{second law in second order}), one can immediately deduce that in the case of vacuum perturbations, the second law of black hole thermodynamics holds at second order if and only if the modified canonical energy flux $\boldsymbol{e}_G$ is non-negative everywhere on the horizon.
	
Next we are going to show that $\boldsymbol{e}_G$ is quadratic in $\delta g_{ab}$ and does not depend on $\delta^2 g_{ab}$. Again, we formally write  $\boldsymbol{\theta}(g,\delta g)=\boldsymbol{D}\delta g$ and $\boldsymbol{B}_\mathcal{H}(g)=\boldsymbol{C} \mathcal{L}_\xi g$ according to (\ref{definition of B}), then $\boldsymbol{\theta} \overset{\mathcal{H}^+}{=} \delta\boldsymbol{B}_\mathcal{H}$ implies that $\boldsymbol{D} \overset{\mathcal{H}^+}{=} \boldsymbol{C}\mathcal{L}_\xi$. Thus
\begin{equation}\label{quantities for calculate EG}
    \begin{split}
        \boldsymbol{\omega} (g;\delta g, \mathcal{L}_\xi\delta g) & = \delta\boldsymbol{\theta}(g; \mathcal{L}_\xi\delta g) - \mathcal{L}_\xi\delta \boldsymbol{\theta} (g; \delta g)=\delta\boldsymbol{D}(\mathcal{L}_\xi\delta g)-(\mathcal{L}_\xi \delta\boldsymbol{D})\delta g,\\
		\delta\boldsymbol{\theta}(g; \delta g) & =\delta\boldsymbol{D}\delta g+\boldsymbol{D}\delta^2 g \overset{\mathcal{H}^+}{=}\delta\boldsymbol{D}\delta g+\boldsymbol{C}\mathcal{L}_\xi\delta^2 g,\\
		\delta^2\boldsymbol{B}_\mathcal{H} & = \delta(\delta\boldsymbol{C}\mathcal{L}_\xi g+\boldsymbol{C}\mathcal{L}_\xi\delta g)=2\delta\boldsymbol{C}\mathcal{L}_\xi\delta g+\boldsymbol{C}\mathcal{L}_\xi\delta^2 g.
    \end{split}
\end{equation}
    
Notice that for an $(n-1)$-form $\boldsymbol{p}$ the pullback of $\xi\cdot \mathrm{d}\boldsymbol{p}$ to the horizon vanishes. By Cartan's magic formula
\begin{equation}
    \mathrm{d}(\xi\cdot\boldsymbol{p}) \overset{\mathcal{H}^+}{=} \mathrm{d}(\xi\cdot\boldsymbol{p})+\xi\cdot \mathrm{d}\boldsymbol{p}=\mathcal{L}_\xi\boldsymbol{p},
\end{equation}
we have
\begin{equation}\label{calculation of EG}
    \begin{split}
		\boldsymbol{e}_G(g; \delta g, \delta g) & \equiv \boldsymbol{\omega} (g;\delta g, \mathcal{L}_\xi\delta g)+\mathrm{d}[\xi \cdot \delta \boldsymbol{\theta}(g,\delta g)-\xi \cdot \delta^2\boldsymbol{B}_\mathcal{H}(g)]\\
		& \overset{\mathcal{H}^+}{=} \delta\boldsymbol{D}(\mathcal{L}_\xi\delta g)-(\mathcal{L}_\xi \delta\boldsymbol{D})\delta g+\mathcal{L}_\xi(\delta\boldsymbol{D}\delta g-2\delta\boldsymbol{C}\mathcal{L}_\xi\delta g)\\
		& = 2\delta\boldsymbol{D}(\mathcal{L}_\xi \delta g)-\mathcal{L}_\xi(2\delta\boldsymbol{C}\mathcal{L}_\xi\delta g).
    \end{split}
\end{equation}
Thus $\boldsymbol{e}_G$ actually does not depend on $\delta^2 g_{ab}$, and to calculate $\boldsymbol{e}_G$ in practice, we need only to calculate $\delta\boldsymbol{\theta}(g;\mathcal{L}_\xi\delta g)$ and $\delta^2 \boldsymbol{B}_\mathcal{H}$ and take their parts which are quadratic in $\delta g_{ab}$.
	

Consider the $f(R)$ gravity, whose Lagrangian is given by 
\begin{equation}
    \boldsymbol{L}_{a_1\cdots a_n}= \frac{1}{16 \pi G_D} f(R) \boldsymbol{\epsilon}_{a_1\cdots a_n}.
\end{equation}
The equation of motion is
\begin{equation}
    E_{ab}=\frac{1}{16 \pi G_D}\left(\frac{1}{2}g_{ab}f-\frac{\partial f}{\partial R}R_{ab}\right.
    +\left.\nabla_a\nabla_b\frac{\partial f}{\partial R}-g_{ab}\square\frac{\partial f}{\partial R} \right)=0,
\end{equation}
where $\square=g^{ab}\nabla_a\nabla_b$. The symplectic potential is
\begin{equation}
    \boldsymbol{\theta}(g;\delta g)_{a_1\cdots a_{n-1}}=\frac{1}{16 \pi G_D}(g^{mc}g^{bd}-g^{md}g^{bc})(\frac{\partial f}{\partial R}\nabla_d-\nabla_d\frac{\partial f}{\partial R})\delta g_{bc} \boldsymbol{\epsilon}_{ma_1\cdots a_{n-1}}
\end{equation}
and
\begin{equation}
    \boldsymbol{B}_{\mathcal{H}a_1\cdots a_{n-1}}\overset{\mathcal{H}^+}{=}- \frac{1}{16 \pi G_D}\xi_m N_d(g^{mc}g^{bd}-g^{md}g^{bc})\frac{\partial f}{\partial R}\mathcal{L}_\xi g_{bc} \boldsymbol{\epsilon}^{(n-1)}.
\end{equation}
According to the aforementioned argument, since (omit the terms containing $\delta^2 g_{ab}$)
\begin{equation}
    \begin{split}
		\delta\boldsymbol{\theta}(g; \mathcal{L}_\xi\delta g)\overset{\mathcal{H}^+}{=} \frac{1}{16 \pi G_D}&\left[\left(\frac{1}{2}f^\prime g^{ef}\delta g_{ef}+f^{\prime\prime}\delta R\right)g^{bc}\mathcal{L}_\xi \mathcal{L}_\xi \delta g_{bc}-f^\prime g^{be}g^{cf}\delta g_{ef}\mathcal{L}_\xi \mathcal{L}_\xi \delta g_{bc}\right.\\
		& \left.-\frac{1}{2}f^\prime g^{be}g^{cf}\mathcal{L}_\xi\delta g_{bc}\mathcal{L}_\xi \delta g_{ef}-f^{\prime\prime}\mathcal{L}_\xi \delta R g^{bc} \mathcal{L}_\xi \delta g_{bc} \right] \boldsymbol{\epsilon}^{(n-1)},
    \end{split}
\end{equation}   
and (also omit the terms containing $\delta^2 g_{ab}$)
\begin{equation}	 
    \delta^2\boldsymbol{B}_\mathcal{H}
    \overset{\mathcal{H}^+}{=} \frac{1}{16 \pi G_D}\left(f^\prime g^{ef}\delta g_{ef} g^{bc}\mathcal{L}_\xi \delta g_{bc}-2f^\prime g^{be}g^{cf}\delta g_{ef}\mathcal{L}_\xi\delta g_{bc}+2f^{\prime\prime}\delta R g^{bc}\mathcal{L}_\xi\delta g_{bc}\right)\boldsymbol{\epsilon}^{(n-1)},
\end{equation}
therefore the modified canonical energy flux is given by
\begin{equation}
    \boldsymbol{e}_G \overset{\mathcal{H}^+}{=} \frac{1}{4 \pi G_D}(\kappa v)^2 \left[f^\prime(R)\left(\delta\sigma_{ab}\delta\sigma^{ab}-\frac{D-3}{D-2}(\delta\theta)^2\right)-2f^{\prime\prime}(R)\mathcal{L}_k\delta R\delta\theta\right]\boldsymbol{\epsilon}^{(n-1)},
\end{equation}        
where $v$ is the affine parameter of the null generator $k^a$ of the future horizon, $\sigma_{ab}=\frac{1}{2}\mathcal{L}_k\gamma_{ab}-\frac{1}{D-2}\theta\gamma_{ab}$ is the shear and $\theta=\partial_v\ln\sqrt{\gamma}$ is the expansion of the generators of the horizon with respect to $v$ as mentioned before.
	
For the vacuum perturbation, the linearized equation of motion is satisfied $\delta E_{ab}=0$, so that on the entire horizon, we have
\begin{equation}
    0=\delta E_{ab}\xi^a\xi^b\overset{\mathcal{H}^+}{=} \frac{\left(\kappa v\right)^2}{16 \pi G_D}\frac{\mathrm{d}}{\mathrm{d}v}\left(f^\prime\delta\theta+f^{\prime\prime}\mathcal{L}_k\delta R\right).
\end{equation}
Suppose that at the late time limit, the spacetime turns to be stationary, i.e. $\lim_{v\rightarrow\infty}(f^\prime\delta\theta+f^{\prime\prime}\mathcal{L}_k\delta R)=0$, then with the above linearized equation of motion, we get 
\begin{equation}
    f^\prime\delta\theta +f^{\prime\prime}\mathcal{L}_k\delta R\overset{\mathcal{H}^+}{=}0.
\end{equation}
Finally, we get the modified canonical energy flux for $f(R)$ gravity
\begin{equation}
    \boldsymbol{e}_G \overset{\mathcal{H}^+}{=} \frac{1}{4 \pi G_D}\left(\kappa v\right)^2 f^\prime(R)\left[\delta\sigma_{ab}\delta\sigma^{ab}+\frac{D-1}{D-2}(\delta\theta)^2\right]\boldsymbol{\epsilon}^{(n-1)},
\end{equation}
and so that
\begin{equation}\label{MCEF}
    \begin{split}
		\frac{\kappa}{2\pi}\Delta\delta^2 S_{\text{dyn}}&=\frac{1}{4 \pi G_D}\int_{\tau_1}^{\tau_2}\mathrm{d} \tau \int_{C(\tau)}dA\left(\kappa v\right)^2 f^\prime(R)\left[\delta\sigma_{ab}\delta\sigma^{ab}+\frac{D-1}{D-2}(\delta\theta)^2\right]\\
		&=\frac{1}{4 \pi G_D}\int_{v_1}^{v_2}dv\int_{C(v)}\mathrm{d}A \ \kappa v f^\prime(R)\left[\delta\sigma_{ab}\delta\sigma^{ab}+\frac{D-1}{D-2}(\delta\theta)^2\right].
    \end{split}
\end{equation}
Where $\tau$ is the Killing parameter for the null Killing generator $\xi^a$, and the Killing parameter $\tau$ is related to the affine parameter $v$ by $v= \frac{C}{\kappa}e^{\kappa \tau}$, which implies that $\mathrm{d}v=\kappa v\mathrm{d}\tau$. (\ref{MCEF}) gives the same result as (\ref{MCEF for FR}), and this provides a nontrivial check for our conclusion obtained via the Einstein frame.

\bibliography{ref}
    
\end{document}